\documentclass[12pt, draftclsnofoot, onecolumn]{IEEEtran}
\usepackage{epsfig,graphicx,subfigure,psfrag,amsmath,cases,bm}
\usepackage{latexsym,amssymb,amsmath,epsfig,subfigure,algorithm,mathtools}
\usepackage{algorithmic}
\usepackage{color}
\usepackage{url}
\usepackage{scrtime}
\usepackage{stfloats}
\usepackage{tablefootnote}
\usepackage{geometry}
\allowdisplaybreaks

\title{Resource Allocation for IRS-assisted Full-Duplex Cognitive Radio Systems\vspace*{4mm}}

\author{\IEEEauthorblockN {Dongfang Xu, \textit{Student Member}, \textit{IEEE}, Xianghao Yu, \textit{Member}, \textit{IEEE},
Yan Sun, \textit{Member}, \textit{IEEE}, Derrick Wing Kwan Ng, \textit{Senior Member}, \textit{IEEE},
and Robert Schober, \textit{Fellow}, \textit{IEEE}\thanks{This work was submitted for presentation in part to the 21st IEEE International Workshop on Signal Processing Advances in Wireless Communications (SPAWC), Atlanta, Georgia, USA, May 2020 \cite{xu2020resource}.}\\
\thanks{
Dongfang Xu, Xianghao Yu, and Robert Schober are with the Institute for Digital Communications, Friedrich-Alexander-University Erlangen-N\"urnberg (FAU), Germany (email:\{dongfang.xu, xianghao.yu, robert.schober\}@fau.de). Yan Sun is with Huawei Technologies Co. Ltd., Shanghai, China (e-mail: sunyan49@huawei.com). Derrick Wing Kwan Ng is with the School of Electrical Engineering and Telecommunications, the University of New South Wales, Australia (email: w.k.ng@unsw.edu.au). 
}}}

\newtheorem{T-Prob}{Transformed Problem}

\DeclareMathOperator{\maxo}{maximize}
\DeclareMathOperator{\mino}{minimize}

 \newcommand{\qed}{\hfill \ensuremath{\blacksquare}}

\begin{document}
\maketitle
\begin{abstract}
In this paper, we investigate the resource allocation design for intelligent reflecting surface (IRS)-assisted full-duplex (FD) cognitive radio systems. In particular, a secondary network employs an FD base station (BS) for serving multiple half-duplex downlink (DL) and uplink (UL) users simultaneously. An IRS is deployed to enhance the performance of the secondary network while helping to mitigate the interference caused to the primary users (PUs). The DL transmit beamforming vectors and the UL receive beamforming vectors at the FD BS, the transmit power of the UL users, and the phase shift matrix at the IRS are jointly optimized for maximization of the total sum rate of the secondary system. The design task is formulated as a non-convex optimization problem taking into account the imperfect knowledge of the PUs' channel state information (CSI) and their maximum interference tolerance. Since the maximum interference tolerance constraint is intractable, we apply a safe approximation to transform it into a convex constraint. To efficiently handle the resulting approximated optimization problem, which is still non-convex, we develop an iterative block coordinate descent (BCD)-based algorithm. This algorithm exploits semidefinite relaxation, a penalty method, and successive convex approximation and is guaranteed to converge to a stationary point of the approximated optimization problem. Our simulation results do not only reveal that the proposed scheme yields a substantially higher system sum rate for the secondary system than several baseline schemes, but also confirm its robustness against CSI uncertainty. Besides, our results illustrate the tremendous potential of IRS for managing the various types of interference arising in FD cognitive radio networks.
\end{abstract}
\vspace*{6mm}
\section{Introduction}
Radio spectrum is a naturally limited resource in wireless communication systems. During the last couple of decades, most of the available spectrum has been licensed and allocated to provide various high data-rate communication services. This has led to the problem of a spectrum crunch for future wireless communication systems \cite{wong2017key}. However, according to measurements of the actual spectrum utilization, e.g. \cite{spectrumreport}, \cite{datla2009spectrum}, a large amount of the allocated spectrum is highly underutilized. As a remedy to improve the spectral efficiency, communication systems employing cognitive radio (CR) technologies have emerged as a promising paradigm to provide communication services to unlicensed secondary systems. One promising approach to spectrum sharing is underlay CR where the secondary system is allowed to use the spectrum concurrently with the primary users (PUs) as long as the quality-of-service (QoS) of the PUs is not severely impaired. Thus, to limit the performance degradation caused to the primary network, the secondary system has to be carefully designed \cite{datla2009spectrum}\nocite{4570202}--\cite{7111366}. For example, the authors of \cite{4570202} developed a joint transmit power allocation and receive beamforming design to minimize the total transmit power of the secondary transmitter, while constraining the interference to the PUs to be below a given threshold. In \cite{7111366}, the authors proposed a multi-objective optimization framework and developed a Pareto-optimal resource allocation algorithm to realize simultaneous wireless power and secure information transfer in CR networks. However, since the CR networks in \cite{4570202}, \cite{7111366} only employ half-duplex (HD) base stations (BSs) and the uplink (UL) and downlink (DL) transmissions are performed in orthogonal frequency bands, the radio spectral resources are still underutilized.
\par
To boost wireless spectral efficiency, full-duplex (FD) communication has recently drawn considerable research interest \cite{6832464}\nocite{afifi2014incorporating,7842235,7760353}--\cite{6493535}. In fact, by incorporating FD BSs into CR networks, the spectral efficiency can be potentially doubled compared to traditional HD CR networks. However, since in CR networks the secondary system has to share the spectrum with the primary system, the QoS of the PUs is inevitably impaired by the simultaneous UL and DL transmissions of the secondary system. In general, compared to the PUs in conventional HD CR networks, because of the larger number of concurrent transmissions, the PUs in FD CR networks suffer from more severe interference, which degrades the performance of the primary network \cite{6832464}, \cite{afifi2014incorporating}. Moreover, the self-interference (SI) and co-channel interference (CCI) caused by the simultaneous DL and UL transmissions, if left unattended, can also significantly degrade the performance of the secondary system \cite{6832464}. To effectively manage the interference in FD CR networks, different resource allocation designs were developed in \cite{7842235}\nocite{7760353}--\cite{6493535}. In \cite{7842235}, the sub-channel assignment, user pairing, and power allocation was jointly optimized to improve the spectral efficiency of a FD CR system. In \cite{7760353}, the authors investigated robust DL beamforming and UL power allocation for minimization of the maximum interference leakage to the PUs taking into account the QoS requirements of the SUs. In \cite{6493535}, multi-antenna precoding and relaying strategies for cooperative FD CR systems were developed to maximize the sum rate of the secondary system taking into account a minimum required data rate for the PUs. Despite these promising results, the PUs in FD CR systems may still suffer from significant interference as the radio frequency (RF) propagation environment of wireless systems is essentially random and largely uncontrollable. In fact, in unfavorable radio propagation environments, the designs proposed in \cite{7842235}\nocite{7760353}--\cite{6493535} cannot mitigate the interference caused to the PUs such that their QoS requirements may be violated. In this case, since the PUs have a higher priority for utilizing the spectrum, the communication in the secondary network may be strictly limited leading to a severe performance degradation of the secondary network. To overcome this problem, more effective interference management methods are urgently needed to facilitate reliable and spectrum-efficient FD CR networks.
\par
Recently, intelligent reflecting surfaces (IRSs) have emerged as a promising solution for harnessing interference in wireless communication systems \cite{di2019smart}\nocite{wu2019intelligent,yu2019miso,xu2019resource,pan2019intelligent,zhang2019capacity,zhang2019multiple}--\cite{8910627}. In particular, an IRS is a planar metasurface comprising a set of small passive low-cost elements, such as phase shifters and printed dipoles, which can be tuned individually to reflect the incident signals with a desired phase shift \cite{cui2014coding}. By adaptively and smartly tuning the phase shifts of the IRS elements according to the dynamic radio propagation environment, the wireless channel can be proactively manipulated, which introduces additional degrees of freedom (DoFs) for resource allocation \cite{di2019smart}. Moreover, the reflected signals can be combined with the non-reflected signals in a constructive or destructive manner to enhance the desired signal power strength or to suppress detrimental interference, which improves the overall system performance. Besides, due to their relatively simple structure \cite{cui2014coding}, IRSs can be flexibly installed on building facades and interior walls, and thus can be smoothly integrated into existing cellular communication systems \cite{di2019smart}. As a result, several works have investigated the application and design of IRS-assisted communication systems \cite{wu2019intelligent}\nocite{yu2019miso}--\cite{xu2019resource}, \cite{yang2019irs}, \cite{yu2019enabling}. For instance, the authors of \cite{wu2019intelligent} considered an IRS-aided multiple-input single-output (MISO) system and studied the joint design of the beamforming at the BS and the IRS to minimize the total BS transmit power. The authors of \cite{yu2019miso} considered an IRS-enhanced single-user system and developed two computationally efficient suboptimal algorithms for maximizing the received power of the user. The authors of \cite{xu2019resource} introduced artificial noise (AN) to improve the physical layer security of an IRS-assisted multiuser communication system and jointly optimized the IRS phase shifts, DL beamformers, and AN design. The authors of \cite{yang2019irs} studied the benefits of IRSs in orthogonal frequency division multiple access systems and formulated a joint transmit power and IRS phase shift optimization problem for maximization of the system throughput. The authors of \cite{yu2019enabling} investigated the joint BS beamforming and IRS phase shift design and proposed two suboptimal algorithms to guarantee physical layer security in an IRS-assisted MISO system. However, the authors of \cite{wu2019intelligent}\nocite{yu2019miso}--\cite{xu2019resource}, \cite{yang2019irs}, \cite{yu2019enabling} considered HD systems, which cannot exploit the full potential of IRSs. In fact, since IRSs naturally operate in a FD manner \cite{8910627}, they can be conveniently incorporated into existing FD CR network concepts to further increase spectrum efficiency. Yet, the designs proposed in \cite{wu2019intelligent}\nocite{yu2019miso}--\cite{xu2019resource}, \cite{yang2019irs}, \cite{yu2019enabling} are not directly applicable to IRS-assisted FD CR networks. In particular, the simultaneous UL and DL transmissions of the secondary system, the superposition of the direct and reflected paths, and the coupling between the DL beamforming vectors, UL transmit powers, and IRS phase shifts makes the resource allocation design for IRS-assisted FD CR networks very challenging. To the best of the authors' knowledge, the design of spectrum-efficient IRS-assisted FD CR networks has not been investigated in the literature, yet. 
\par
Motivated by the above discussion, in this paper, we integrate IRSs into FD CR networks and investigate the corresponding resource allocation algorithm design. In particular, as the secondary system is allowed to share the spectrum of the primary system as long as the QoS of the PUs is not severely compromised, the IRS is utilized to establish a favorable radio propagation environment. In particular, we aim to maximize the sum rate of the secondary system by jointly optimizing the DL transmit beamformers, the UL transmit power, the UL receive beamformers, and the IRS phase shifts. The problem formulation takes into account the imperfect knowledge of the channel state information (CSI) of the PUs at the FD BS of the secondary system and the maximum interference leakage tolerance of the PUs. Since the maximum interference leakage tolerance constraint is intractable, we transform it into a convex constraint by applying a safe approximation. Due to the coupling between the optimization variables and the unit-modulus constraint of the IRS phase shifts, even with the safe approximation, the formulated problem is still highly non-convex and it is very challenging to obtain the optimal solution. Hence, we propose a block coordinate descent (BCD)-based iterative algorithm to obtain a suboptimal solution \cite{tseng2001convergence}. In particular, by applying successive convex approximation (SCA) \cite{dinh2010local} and semidefinite relaxation (SDR), the DL transmit beamforming and UL power allocation policies are obtained with the other optimization variables being fixed. Then, we derive the closed-form optimal solution for the receive beamforming vector of the secondary BS given the other optimization variables. Subsequently, we obtain the phase shift matrix of the IRS by applying a penalty method \cite{ben1997penalty} and SCA. The developed BCD algorithm is guaranteed to converge to a stationary point of the approximated optimization problem. Simulation results reveal that IRSs and the proposed algorithm can significantly enhance the performance of secondary networks while efficiently mitigating the interference to the PUs.
\par
\textit{Notations}: In this paper, boldface lower case and boldface capital letters denote vectors and matrices, respectively. $\mathbb{N}$ denotes the set of nonnegative integers. $\mathbb{R}^{N\times M}$ and $\mathbb{C}^{N\times M}$ denote the space of $N\times M$ real-valued and complex-valued matrices, respectively. $\Re\left \{ \cdot \right \}$ extracts the real part of a complex number. $\mathbb{H}^{N}$ denotes the set of all $N$-dimensional complex Hermitian matrices. $\mathbf{I}_{N}$ indicates the $N\times N$ identity matrix. $|\cdot|$ and $||\cdot||_2$ denote the absolute value of a complex scalar and the $l_2$-norm of a vector, respectively. $\mathbf{A}^T$, $\mathbf{A}^*$, and $\mathbf{A}^H$ stand for the transpose, the conjugate, and the conjugate transpose of matrix $\mathbf{A}$, respectively. $\mathbf{A}\succeq\mathbf{0}$ indicates that $\mathbf{A}$ is a positive semidefinite matrix. $\mathrm{Rank}(\mathbf{A})$, $\mathrm{Tr}(\mathbf{A})$, $\left [ \mathbf{A} \right ]_{i,i}$, and $\left \| \mathbf{A} \right \|_*$ denote the rank, the trace, the $(i,i)$-entry, and the trace norm of matrix $\mathbf{A}$, respectively. $x_i$ denotes the $i$-th element of vector $\mathbf{x}$. $\mathrm{Diag}(\mathbf{X})$ represents a diagonal matrix
whose diagonal elements are extracted from the main diagonal of matrix $\mathbf{X}$; $\mathrm{diag}(\mathbf{x})$ denotes an $N\times N$ diagonal matrix with main diagonal elements $x_1, \cdots, x_N$. $\mathcal{E}\left \{ \cdot \right \}$ denotes statistical expectation. $\sim$ and $\overset{\Delta }{=}$ stand for ``distributed as'' and ``defined as'', respectively. The distribution of a circularly symmetric complex Gaussian random variable with mean $\mu$ and variance $\sigma^2$ is denoted by $\mathcal{CN}(\mu ,\sigma^2)$. The gradient vector of function $f(\mathbf{x})$ with respect to $\mathbf{x}$ is denoted by $\nabla_{\mathbf{x}} f(\mathbf{x})$. $\mathbf{x}^{\dagger}$ denotes the optimal value of optimization variable $\mathbf{x}$.
\section{System Model}
\begin{figure}[t]\vspace*{-2mm}
 \centering
\includegraphics[width=3.8in]{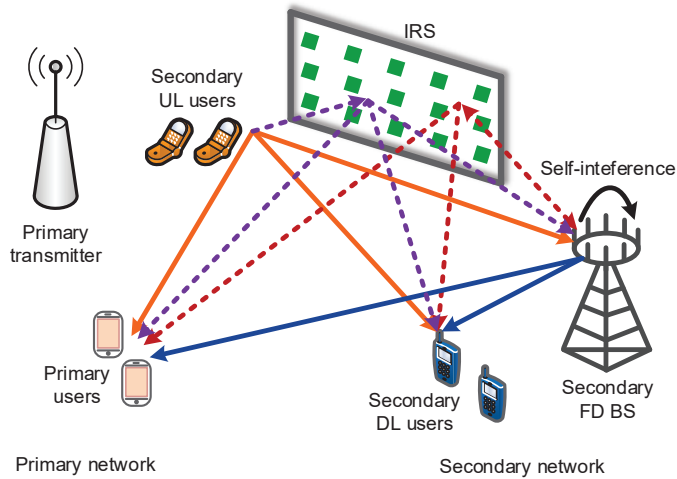} \vspace*{-2mm}
\caption{CR system comprising a secondary FD base station, $J=2$ secondary UL users, and $K=2$ secondary DL users sharing the spectrum with $I=2$ PUs. The IRS is deployed to enhance the system performance of the secondary network and to mitigate the interference to the PUs. The direct paths and reflected paths are denoted by solid arrows and dashed arrows, respectively. The signals of the primary transmitter are not shown for clarity.
}
\label{system_model}\vspace*{-2mm}
\end{figure}
In this section, we present the IRS-assisted multiuser FD CR network model and discuss our assumptions regarding the CSI available for resource allocation.
\subsection{IRS-Assisted Full-Duplex Cognitive Radio System Model}
We consider a narrow-band IRS-assisted CR communication system\footnote{In this paper, we consider an underlay CR network \cite{liu2016nonorthogonal} where the secondary FD BS opportunistically coexists with the
primary transmitter when the interference leakage to the
PUs remains below a certain threshold.} consisting of a primary license-holding network and a secondary unlicensed network, cf. Figure \ref{system_model}. In particular, the primary network comprises one primary transmitter and $I$ PUs, while the secondary network includes one secondary FD BS, $J$ UL users, and $K$ DL users. The primary transmitter, the $I$ PUs, and the $K+J$ secondary users are single-antenna HD devices. The secondary FD BS is equipped with $N_{\mathrm{T}}>1$ antennas\footnote{To facilitate reliable UL signal detection, we assume that the number of antennas equipped at the secondary FD BS is equal to or larger than the number of secondary UL users, i.e., $N_{\mathrm{T}}\geq J$.}, indexed by $\mathcal{N}\overset{\Delta }{=}\left \{1,\cdots ,N_{\mathrm{T}} \right \}$, and simultaneously performs DL transmission and UL reception in the same frequency band\footnote{Simultaneous transmission and reception with the same antenna can be realized by employing a circulator-based FD radio transceiver, as demonstrated in \cite{bharadia2013full}.}. Due to the spectrum sharing, the QoS of the primary network is impaired by interference leakage from the secondary network. To effectively suppress the interference and improve the system performance of the secondary network, an IRS is deployed. In particular, the IRS comprises $M$ phase shifters, indexed by $\mathcal{M}\overset{\Delta }{=}\left \{1,\cdots ,M \right \}$, and is programmable and reconfigurable via an IRS controller. 
For notational simplicity, we define sets $\mathcal{I}=\left \{1,\cdots ,I \right \}$, $\mathcal{J}=\left \{1,\cdots ,J \right \}$, and $\mathcal{K}=\left \{1,\cdots ,K \right \}$ for the indices of the PUs, secondary UL users, and secondary DL users, respectively.
\par
In a given time slot, the secondary FD BS transmits signal $\underset{k\in\mathcal{K}}{\sum }\mathbf{w}_kd^{\mathrm{DL}}_k$ to the $K$ DL users, where $d^{\mathrm{DL}}_k\in\mathbb{C}$ and $\mathbf{w}_k\in \mathbb{C}^{\mathit{N}_{\mathrm{T}}\times 1}$ denote the information symbol for secondary DL user $k$ and the corresponding beamformer, respectively. Without loss of generality, we assume $\mathcal{E}\{\left |d^{\mathrm{DL}}_k \right|^2\}=1$, $\forall\mathit{k} \in \mathcal{K}$. The received signals at PU $i$, the secondary FD BS, and secondary DL user $k$ are given by
\begin{eqnarray}
\label{PUsig}
y_i^{\mathrm{P}}\hspace*{-2mm}&\hspace*{-2mm}=&\hspace*{-3mm}s_i^{\mathrm{P}}+\underset{k\in\mathcal{K} }{\sum }\hspace*{1mm}\mathbf{l}^H_{\mathrm{D},i}\mathbf{w}_kd^{\mathrm{DL}}_k+\underset{k\in\mathcal{K} }{\sum }\hspace*{1mm}\mathbf{l}^H_{\mathrm{R},i}\bm{\Psi} \mathbf{F}\mathbf{w}_kd^{\mathrm{DL}}_k+\underset{j\in\mathcal{J} }{\sum }\sqrt{p_j}e_{i,j}d^{\mathrm{UL}}_j
+\underset{j\in\mathcal{J} }{\sum }\sqrt{p_j}\hspace*{1mm}\mathbf{l}^H_{\mathrm{R},i}\bm{\Psi}\mathbf{h}_{\mathrm{R},j}d^{\mathrm{UL}}_j+n_i^{\mathrm{P}},\\
\label{ULsig}
\mathbf{y}^{\mathrm{UL}}\hspace*{-2mm}&\hspace*{-2mm}=&\hspace*{-3mm}\underbrace{\underset{j\in\mathcal{J}}{\sum }\sqrt{p_j}\hspace*{1mm}\mathbf{h}_{\mathrm{D},j}d^{\mathrm{UL}}_j+\underset{j\in\mathcal{J} }{\sum }\sqrt{p_j}\hspace*{1mm} \mathbf{F}^H\bm{\Psi}\mathbf{h}_{\mathrm{R},j}d^{\mathrm{UL}}_j}_{\text{desired signal}}+\underbrace{\mathbf{S}\underset{k\in\mathcal{K} }{\sum }\mathbf{w}_kd^{\mathrm{DL}}_k+\underset{k\in\mathcal{K} }{\sum }\mathbf{F}^H\bm{\Psi}\mathbf{F}\mathbf{w}_kd^{\mathrm{DL}}_k}_{\text{self-interference}}+\mathbf{n}^{\mathrm{UL}},\\
y_k^{\mathrm{DL}}\hspace*{-2mm}&\hspace*{-2mm}=&\hspace*{-3mm}\underbrace{\mathbf{g}^H_{\mathrm{D},k}\mathbf{w}_kd^{\mathrm{DL}}_k+\mathbf{g}^H_{\mathrm{R},k}\bm{\Psi}\mathbf{F}\mathbf{w}_kd^{\mathrm{DL}}_k}_{\text{desired signal}}+\underbrace{\underset{r\in\mathcal{K}\setminus \left \{ k \right \} }{\sum }\mathbf{g}^H_{\mathrm{D},k}\mathbf{w}_rd^{\mathrm{DL}}_r+\underset{r\in\mathcal{K}\setminus \left \{ k \right \} }{\sum }\mathbf{g}^H_{\mathrm{R},k}\bm{\Psi}\mathbf{F}\mathbf{w}_rd^{\mathrm{DL}}_r}_{\text{multiuser interference}}\notag\\&+&\underbrace{\underset{j\in\mathcal{J} }{\sum }\sqrt{p_j}q_{j,k}d^{\mathrm{UL}}_j+\underset{j\in\mathcal{J} }{\sum }\sqrt{p_j}\mathbf{g}^H_{\mathrm{R},k}\bm{\Psi} \mathbf{h}_{\mathrm{R},j}d^{\mathrm{UL}}_j}_{\text{co-channel interference}}+n_k^{\mathrm{DL}},
\end{eqnarray}
respectively. $s_i^{\mathrm{P}}$ in \eqref{PUsig} denotes the received signal originating from the primary transmitter. $\mathbf{l}_{\mathrm{D},i}\in \mathbb{C}^{\mathit{N}_{\mathrm{T}}\times 1}$ and $\mathbf{l}_{\mathrm{R},i}\in \mathbb{C}^{M\times 1}$ denote the channel vector between the secondary BS and the $i$-th PU and the channel vector between the IRS and the $i$-th PU\footnote{The delays between the signal propagating through the direct path and the reflected path via the IRS are typically much shorter than the signal duration. For instance, for a cell with a radius of 50 meters as considered in our simulations, cf. Figure \ref{simu_model}, the maximum round-trip delay is 0.33 $\mu$s, which is significantly shorter than the 70 $\mu$s symbol
duration in the Long-Term Evolution (LTE) standard \cite{ghosh2010fundamentals}. Thus, we neglect the impact of intersymbol interference in this paper.}, respectively. Diagonal matrix $\mathbf{\Psi}=\mathrm{diag}\left ( e^{j\psi_1}, \cdots, e^{j\psi_M}  \right )$ represents the phase shift matrix of the IRS \cite{wu2019intelligent}, where $\psi_m\in \left [ -\pi ,\pi  \right ]$, $\forall m \in \mathcal{M}$, is the phase shift introduced by the $m$-th IRS element. Matrix $\mathbf{F}\in\mathbb{C}^{\mathit{M}\times\mathit{N}_{\mathrm{T}}}$ models the channel between the secondary FD BS and the IRS. Variables  $d_j^{\mathrm{UL}}\in\mathbb{C}$ and $p_j$ are the data symbol and the corresponding power transmitted by secondary UL user $j$ to the secondary FD BS, respectively. We assume $\mathcal{E}\{\left |d _{j}^{\mathrm{UL}} \right|^2\}=1$ without loss of generality. The channel gain between secondary UL user $j$ and PU $i$ is denoted by $e_{i,j}$. $\mathbf{h}_{\mathrm{D},j}\in \mathbb{C}^{\mathit{N}_{\mathrm{T}}\times 1}$ and $\mathbf{h}_{\mathrm{R},j}\in \mathbb{C}^{M\times 1}$ denote the channel vector between the secondary BS and secondary UL user $j$ and the channel vector between the IRS and secondary UL user $j$, respectively. $\mathbf{S}\underset{k\in\mathcal{K} }{\sum }\mathbf{w}_kd^{\mathrm{DL}}_k$ in \eqref{ULsig} represents the SI resulting from the DL transmission with $\mathbf{S}\in\mathbb{C}^{\mathit{N}_{\mathrm{T}}\times\mathit{N}_{\mathrm{T}}}$ denoting the SI channel matrix of the secondary FD BS. The term $\underset{k\in\mathcal{K} }{\sum }\mathbf{F}^H\bm{\Psi}\mathbf{F}\mathbf{w}_kd^{\mathrm{DL}}_k$ in \eqref{ULsig} denotes the SI introduced by the reflection of the DL transmit signal by the IRS. $\mathbf{g}_{\mathrm{D},k}\in \mathbb{C}^{\mathit{N}_{\mathrm{T}}\times 1}$ and $\mathbf{g}_{\mathrm{R},k}\in \mathbb{C}^{M\times 1}$ denote the channel vector between the secondary BS and the $k$-th DL user and the channel vector between the IRS and the $k$-th DL user, respectively. The channel gain between secondary UL user $j$ and secondary DL user $k$ is denoted by $q_{j,k}$. $\mathbf{n}^{\mathrm{UL}}\sim\mathcal{CN}(\mathbf{0},\sigma_{\mathrm{U}}^2\mathbf{I}_{N_{\mathrm{T}}})$ and $n_k^{\mathrm{DL}}\sim\mathcal{CN}(0,\sigma_{\mathrm{n}_k}^2)$ denote the equivalent additive white Gaussian
noises (AWGNs) at the secondary FD BS and secondary DL user $k$, which capture the combined effect of thermal noise, signal processing noise, and the interference stemming from the primary network \cite{zhang2009robust}. $n_i^{\mathrm{P}}$ includes the
joint effects of thermal noise and signal processing noise
at PU $i$.
\subsection{Channel State Information}
In this paper, we focus on slowly time-varying channels. During the channel estimation phase of the secondary network, the secondary FD BS can reliably estimate all links of the secondary network with the assistance of the SUs and the IRS \cite{wang2019channel}. As a result, we assume that the perfect CSI of the secondary network is available at the secondary FD BS for resource allocation. However, this assumption may not be valid for the channels between the secondary network and the PUs. In practice, the PUs can not be expected to directly interact with the secondary FD BS. Moreover, the PUs may be idle for a long period of time due to bursty data transmission. As a result, the CSI of the PUs can be obtained only occasionally at the secondary FD BS when the PUs are active in the primary network, which leads to outdated PU CSI at the FD BS. In this paper, we develop a worst-case optimization framework and adopt a deterministic model to capture the impact of imperfect PU CSI on resource allocation design \cite{wang2009worst}. Specifically, the CSI of the link between the FD BS and PU $i$, i.e., $\mathbf{l}_{\mathrm{D},i}$, the CSI of the link between the IRS and PU $i$, i.e., $\mathbf{l}_{\mathrm{R},i}$, and the CSI of the link between PU $i$ and secondary UL user $j$ are modeled as:
\begin{eqnarray}
    &&\mathbf{l}_{\mathrm{D},i}=\overline{\mathbf{l}}_{\mathrm{D},i}+\Delta \mathbf{l}_{\mathrm{D},i}~~\mbox{and}~~
    \Omega_{\mathrm{D},i}\overset{\Delta }{=}\left \{ \mathbf{l}_{\mathrm{D},i}\hspace*{1mm}|\hspace*{1mm}\Delta \mathbf{l}_{\mathrm{D},i}^H\Delta \mathbf{l}_{\mathrm{D},i}\leq \varepsilon_{\mathrm{D},i} ^2 \right \},\\
    &&\mathbf{l}_{\mathrm{R},i}=\overline{\mathbf{l}}_{\mathrm{R},i}+\Delta \mathbf{l}_{\mathrm{R},i}~~\mbox{and}~~
    \Omega_{\mathrm{R},i}\overset{\Delta }{=}\left \{ \mathbf{l}_{\mathrm{R},i}\hspace*{1mm}|\hspace*{1mm}\Delta \mathbf{l}_{\mathrm{R},i}^H\Delta \mathbf{l}_{\mathrm{R},i}\leq \varepsilon_{\mathrm{R},i} ^2 \right \},\\
    &&e_{i,j}=\overline{e}_{i,j}+\Delta e_{i,j}~~\mbox{and}~~
    \Omega_{i,j}\overset{\Delta }{=}\left \{ e_{i,j}\hspace*{1mm}|\hspace*{1mm}\Delta e_{i,j}^H\Delta e_{i,j}\leq \varepsilon_{i,j} ^2 \right \},
\end{eqnarray}
respectively, where $\overline{\mathbf{l}}_{\mathrm{D},i}$, $\overline{\mathbf{l}}_{\mathrm{R},i}$, and $\overline{e}_{i,j}$ are the CSI estimates and $\Delta \mathbf{l}_{\mathrm{D},i}$, $\Delta \mathbf{l}_{\mathrm{R},i}$, and $\Delta e_{i,j}$ are the corresponding unknown estimation errors, respectively. The continuous sets $\Omega_{\mathrm{D},i}$, $\Omega_{\mathrm{R},i}$, and $\Omega_{i,j}$ contain all possible channel estimation errors with bounded magnitude $\varepsilon_{\mathrm{D},i}$, $\varepsilon_{\mathrm{R},i}$, and $\varepsilon_{i,j}$, respectively.

\section{Resource Allocation Problem Formulation}
In this section, after introducing the adopted performance metrics, we formulate the proposed resource allocation optimization problem.
\subsection{Performance Metrics}
The achievable rate (bits/s/Hz) of secondary DL user $k$ is given by $R_k^{\mathrm{DL}}=\mathrm{log}_2(1+\Gamma_k^{\mathrm{DL}})$, where $\Gamma_k^{\mathrm{DL}}$
is the receive signal-to-noise-plus-interference ratio (SINR) of secondary DL user $k$ and given by
\begin{equation}
\label{downlinkR}
\Gamma_k^{\mathrm{DL}}=\frac{\left | \mathbf{g}_{\mathrm{D},k}^H\mathbf{w}_k+\mathbf{g}_{\mathrm{R},k}^H \mathbf{\Psi}\mathbf{F}\mathbf{w}_k\right |^2}{\underset{r\in\mathcal{K}\setminus \left \{ k \right \} }{\sum} \left |\mathbf{g}_{\mathrm{D},k}^H\mathbf{w}_r+\mathbf{g}_{\mathrm{R},k}^H \mathbf{\Psi}\mathbf{F}\mathbf{w}_r\right |^2+\underset{j\in\mathcal{J}}{\sum}p_j \left | q_{j,k}+ \mathbf{g}_{\mathrm{R},k}^H \mathbf{\Psi}\mathbf{h}_{\mathrm{R},j}\right |^2 +\sigma^2_{\mathrm{n}_k}}.
\end{equation}
On the other hand, the achievable rate (bits/s/Hz) of secondary UL user $j$ is given by $R_j^{\mathrm{UL}}=\mathrm{log}_2(1+\Gamma_j^{\mathrm{UL}})$, where $\Gamma_j^{\mathrm{UL}}$
is the receive SINR of secondary UL user $j$ and given by
\begin{equation}
\label{uplinkR}
    \Gamma_j^{\mathrm{UL}}=\frac{p_j\left | \mathbf{v}_j^H\mathbf{h}_{\mathrm{D},j}+\mathbf{v}_j^H\mathbf{F}^H\mathbf{\Psi}\mathbf{h}_{\mathrm{R},j}\right |^2}{\underset{t\in\mathcal{J}\setminus \left \{ j \right \} }{\sum} p_t\left |\mathbf{v}_j^H\mathbf{h}_{\mathrm{D},t}+\mathbf{v}_j^H\mathbf{F}^H\mathbf{\Psi}\mathbf{h}_{\mathrm{R},t}\right |^2+D_j^{\mathrm{SI}}+\sigma^2_{\mathrm{U}}\left \| \mathbf{v}_j \right \|^2}.
\end{equation}
Here, $\mathbf{v}_j\in\mathbb{C}^{N_{\mathrm{T}}\times 1}$ is the receive beamforming vector for decoding the message of secondary UL user $j$ at the secondary FD BS. We note that due to the limited dynamic range of the receiver, the SI cannot be suppressed completely even if perfect CSI of the SI channel is available at the secondary FD BS \cite{day2012full}. Thus, similar to \cite{day2012full}, \cite{7463025}, we model the residual SI after cancellation at each receive antenna as an independent Gaussian distortion noise with zero mean and a variance proportional to the power received at that antenna. In particular, the term $D_j^{\mathrm{SI}}$ in \eqref{uplinkR} is given by \cite[Eq. (4)]{day2012full}
\begin{equation}
\label{SIsignal}
      D_j^{\mathrm{SI}}\hspace{-1mm}=\hspace{-1mm}\mathrm{Tr}\Big (\eta\mathbf{v}_j\mathbf{v}_j^H\mathrm{Diag}\big (\underset{k\in\mathcal{K}}{\sum }\underbrace{\mathbf{S}\mathbf{w}_k\mathbf{w}_k^H\mathbf{S}^H\hspace{-1mm}}_{\text{self-interference}}+\hspace{-1mm}\underbrace{\mathbf{F}^H\hspace{-0.5mm}\bm{\Psi}\mathbf{F}\mathbf{w}_k\mathbf{w}_k^H\mathbf{F}^H\hspace{-0.5mm}\bm{\Psi}^H\hspace{-0.5mm}\mathbf{F}\hspace{-1mm}+\hspace{-1mm}\mathbf{F}^H\hspace{-0.5mm}\bm{\Psi}\mathbf{F}\mathbf{w}_k\mathbf{w}_k^H\mathbf{S}^H\hspace{-1mm}+\hspace{-1mm}\mathbf{S}\mathbf{w}_k\mathbf{w}_k^H\mathbf{F}^H\hspace{-0.5mm}\bm{\Psi}^H\hspace{-0.5mm}\mathbf{F}}_{\text{reflected interference}}\big )\hspace{-1mm}\Big ),
\end{equation}
where the constant $\eta$, $0<\eta\ll1$, captures the impact of the residual interference after SI cancellation at the secondary FD BS \cite{masmoudi2016channel}. We note that due to the propagation attenuation between the FD BS and the IRS, the reflected interference in \eqref{SIsignal} is negligible \footnote{For a CR network where the IRS is 100 meter away from the FD BS and a path loss exponent of 2, the term $\mathbf{F}^H\bm{\Psi}\mathbf{F}\mathbf{w}_k\mathbf{w}_k^H\mathbf{F}^H\bm{\Psi}^H\mathbf{F}+\mathbf{F}^H\bm{\Psi}\mathbf{F}\mathbf{w}_k\mathbf{w}_k^H\mathbf{S}^H+\mathbf{S}\mathbf{w}_k\mathbf{w}_k^H\mathbf{F}^H\bm{\Psi}^H\mathbf{F}$ is attenuated by approximately a factor of $10^{-8}$ compared to the term $\mathbf{S}\mathbf{w}_k\mathbf{w}_k^H\mathbf{S}^H$.} compared to self-interference. As a result, we can approximate \eqref{SIsignal} as follows
\begin{equation}
\label{SIsignal2}
      D_j^{\mathrm{SI}}\approx\mathrm{Tr}\Big (\eta\mathbf{v}_j\mathbf{v}_j^H\mathrm{Diag}\big (\underset{k\in\mathcal{K}}{\sum }\mathbf{S}\mathbf{w}_k\mathbf{w}_k^H\mathbf{S}^H\big )\Big ).
\end{equation}
\subsection{Optimization Problem Formulation}
In this paper, we optimize $\mathbf{w}_k$, $\mathbf{v}_j$, $p_j$, and $\mathbf{\Psi}$ to maximize the system sum rate of the secondary network while limiting the interference caused by the secondary network to the PUs. The corresponding optimization problem is formulated as follows
\vspace*{-1mm}
\begin{eqnarray}
\label{prob1}
&&\hspace*{2mm}\underset{\mathbf{w}_k,\mathbf{v}_j,p_j,\mathbf{\Psi}}{\maxo} \,\, \,\, F\big(\mathbf{w}_k,\mathbf{v}_j,p_j,\mathbf{\Psi}\big)\overset{\Delta }{=}\underset{ j\in\mathcal{J}}{\sum}\omega_j^{\mathrm{UL}}\mathrm{log}_2\left(1+\Gamma_j^{\mathrm{UL}}\right)+\underset{ k\in\mathcal{K}}{\sum}\omega_k^{\mathrm{DL}}\mathrm{log}_2(1+\Gamma_k^{\mathrm{DL}})\\
\mbox{s.t.}\hspace*{-4mm}
&&\mbox{C1:~}\underset{k\in\mathcal{K}}{\sum }\left \| \mathbf{w}_k \right \|^2\leq P_{\mathrm{max}}^{\mathrm{DL}},\hspace*{6mm}\mbox{C2:~}0\leq p_j\leq p_{j,\mathrm{max}},~\forall j,\hspace*{6mm}\mbox{C3:~}\left |  \left [ \mathbf{\Psi} \right ]_{m,m}\right |=1,~\forall m,\notag\\
&&\mbox{C4:~}\underset{\substack{\mathbf{l}_{\mathrm{D},i}\in\Omega_{\mathrm{D},i}\\\mathbf{l}_{\mathrm{R},i}\in\Omega_{\mathrm{R},i}\\\mathbf{e}_{i,j}\in\Omega_{i,j}}}{\mathrm{max}}\hspace*{2mm}\underset{k\in \mathcal{K} }{\sum }\left | \mathbf{l}_{\mathrm{D},i}^H\mathbf{w}_k +\mathbf{l}_{\mathrm{R},i}^H\mathbf{\Psi }\mathbf{F}\mathbf{w}_k \right |^2+\underset{j\in \mathcal{J} }{\sum }p_j\left | e_{i,j}+ \mathbf{l}_{\mathrm{R},i}^H\mathbf{\Psi }\mathbf{h}_{\mathrm{R},j} \right |^2\leq
p_{\mathrm{tol}_i},~\forall i\notag,
\end{eqnarray}
where $\omega_j^{\mathrm{UL}}\geq 0$ and $\omega_k^{\mathrm{DL}}\geq 0$ denote predefined weights for secondary UL user $j$ and DL user $k$, which can be used to prioritize the UL and DL users. $P_{\mathrm{max}}^{\mathrm{DL}}>0$ and $p_{j,\mathrm{max}}>0$ in constraints C1 and C2 limit the maximum transmit powers of the secondary FD BS and secondary UL user $j$, respectively. Constraint C3 guarantees that the diagonal phase shift matrix $\mathbf{\Psi}$ has $M$ unit modulus components on its main diagonal. C4 constrains the maximum tolerable interference leakage. In particular, despite the imperfection of the CSI, the secondary network is required to ensure that the interference leakage to PU $i$ does not exceed the maximum interference tolerance $p_{\mathrm{tol}_i}$.
\par
We note that problem \eqref{prob1} is a highly non-convex optimization problem. In particular, the coupling of the optimization variables, the non-convexity of the objective function, the unit-modulus constraint C3, and the semi-infinite constraint C4 are the main obstacles for solving the considered resource allocation problem efficiently. To the best of the authors' knowledge, the globally optimal solution of this problem is in general intractable. In the next section, we develop a suboptimal BCD-based iterative algorithm to solve problem \eqref{prob1} with polynomial time complexity.
\section{Solution of the Optimization Problem}
In this section, we first employ a safe approximation to convert constraint C4 to a set of convex constraints. Then, we propose a BCD-based algorithm to tackle the approximated problem, which is still highly non-convex. In fact, BCD is a widely applicable approach that divides coupled optimization variables into several blocks and solves the optimization problem for one block at a time while fixing the variables in the other blocks \cite{tseng2001convergence}. In particular, we divide the optimization variables into three blocks: $\left \{\mathbf{w}_k, p_j\right \}$, $\left \{\mathbf{v}_j\right \}$, and $\left \{\mathbf{\Psi}\right \}$. By employing SCA and SDR, we obtain the transmit beamforming vector $\mathbf{w}_k$ and transmit power $p_j$. Then, we derive a closed-form solution for receive beamforming vector $\mathbf{v}_j$. Subsequently, we solve for $\mathbf{\Psi}$ by applying a penalty method and SCA.
\subsection{Transformation of the Semi-Infinite Constraints}
In the literature, semi-infinite constraints are commonly transformed into tractable linear matrix inequality (LMI) constraints \cite{wang2009worst}. However, due to the coupling between the optimization variables and the coupling between the signals of the direct and reflect paths, it is challenging to transform constraint C4 into an LMI that is jointly convex with respect to $\mathbf{w}_k$ and $\bm{\Psi}$. To facilitate robust resource allocation algorithm design, we first apply inequality $\left | a+b+c \right |^2\leq 3\left | a \right |^2+3\left | b \right |^2+3\left | c \right |^2$, where $a$, $b$, and $c$ are complex numbers, to the left hand side of constraint C4 to obtain a tractable upper bound\footnote{We note that the upper bound becomes tight when $a$, $b$, and $c$ have similar values.}. In particular, a subset of the set defined by constraint C4 is given by
\begin{eqnarray}
    \overline{\mbox{C4}}\mbox{:}\hspace*{2mm}\underset{\substack{\mathbf{l}_{\mathrm{D},i}\in\Omega_{\mathrm{D},i}\\\mathbf{l}_{\mathrm{R},i}\in\Omega_{\mathrm{R},i}\\\mathbf{e}_{i,j}\in\Omega_{i,j}}}{\mathrm{max}}\hspace*{-4mm}&&\underset{k\in \mathcal{K} }{\sum }\left (\left | \Delta\mathbf{l}_{\mathrm{D},i}^H\mathbf{w}_k \right |^2+\left | \Delta\mathbf{l}_{\mathrm{R},i}^H\mathbf{\Psi }\mathbf{F}\mathbf{w}_k \right |^2+\left | \overline{\mathbf{l}}_{\mathrm{D},i}^H\mathbf{w}_k +\overline{\mathbf{l}}_{\mathrm{R},i}^H\mathbf{\Psi }\mathbf{F}\mathbf{w}_k \right |^2\right )\notag\\
    +\hspace*{-6mm}&&\underset{j\in \mathcal{J} }{\sum }p_j\left ( \left | \Delta e_{i,j} \right |^2+\left |\Delta \mathbf{l}_{\mathrm{R},i}^H\mathbf{\Psi }\mathbf{h}_{\mathrm{R},j} \right |^2+\left | \overline{e}_{i,j}+ \overline{\mathbf{l}}_{\mathrm{R},i}^H\mathbf{\Psi }\mathbf{h}_{\mathrm{R},j} \right |^2\right )\leq
\frac{p_{\mathrm{tol}_i}}{3},~\forall i.
\end{eqnarray}
In the remainder of the paper, we tackle the following approximated optimization problem:
\vspace*{-1mm}
\begin{eqnarray}
\label{altprob1}
&&\hspace*{-14mm}\underset{\mathbf{w}_k,\mathbf{v}_j,p_j,\mathbf{\Psi}}{\maxo} \,\, \,\, F\big(\mathbf{w}_k,\mathbf{v}_j,p_j,\mathbf{\Psi}\big)\\
\mbox{s.t.}\hspace*{-4mm}
&&\mbox{C1},\mbox{C2},\mbox{C3},\overline{\mbox{C4}}\notag.
\end{eqnarray}
We note that any feasible solution of \eqref{altprob1} is also a feasible solution of \eqref{prob1}. Hence, \eqref{altprob1} is a safe approximation of \eqref{altprob1} \cite{li2013safe}. Then, we define slack variables $\beta_i$, $\gamma_i$, and $\tau_i$ and rewrite constraint $\overline{\mbox{C4}}$ equivalently in terms of the following constraints:
\begin{eqnarray}
&&\overline{\mbox{C4}}\mbox{a:~}\underset{e_{i,j}\in\Omega_{i,j}}{\mathrm{max}}\hspace*{2mm}\underset{j\in \mathcal{J} }{\sum }p_j\left | \Delta e_{i,j} \right |^2+\beta_i\leq
\frac{p_{\mathrm{tol}_i}}{3},~\forall i,\\
&&\overline{\mbox{C4}}\mbox{b:~}\underset{\mathbf{l}_{\mathrm{D},i}\in\Omega_{\mathrm{D},i}}{\mathrm{max}}\hspace*{2mm}\underset{k\in \mathcal{K} }{\sum }\left | \Delta\mathbf{l}_{\mathrm{D},i}^H\mathbf{w}_k \right |^2+\gamma_i\leq \beta_i,~\forall i,\\
&&\overline{\mbox{C4}}\mbox{c:~}\underset{\mathbf{l}_{\mathrm{R},i}\in\Omega_{\mathrm{R},i}}{\mathrm{max}}\hspace*{2mm}\left ( \underset{k\in \mathcal{K} }{\sum }\left | \Delta\mathbf{l}_{\mathrm{R},i}^H\mathbf{\Psi }\mathbf{F}\mathbf{w}_k \right |^2+\underset{j\in \mathcal{J} }{\sum }p_j\left |\Delta \mathbf{l}_{\mathrm{R},i}^H\mathbf{\Psi }\mathbf{h}_{\mathrm{R},j} \right |^2\right )+\tau_i\leq \gamma_i,~\forall i,\\
&&\overline{\mbox{C4}}\mbox{d:~}\underset{k\in \mathcal{K} }{\sum }\left | \overline{\mathbf{l}}_{\mathrm{D},i}^H\mathbf{w}_k +\overline{\mathbf{l}}_{\mathrm{R},i}^H\mathbf{\Psi }\mathbf{F}\mathbf{w}_k \right |^2+\underset{j\in \mathcal{J} }{\sum }p_j\left | \overline{e}_{i,j}+ \overline{\mathbf{l}}_{\mathrm{R},i}^H\mathbf{\Psi }\mathbf{h}_{\mathrm{R},j} \right |^2\leq\tau_i,~\forall i.\label{c4d}
\end{eqnarray}
We note that $\overline{\mbox{C4}}\mbox{d}$ is convex in $\mathbf{w}_k$ and $\mathbf{\Psi}$ individually while $\overline{\mbox{C4}}\mbox{a}$, $\overline{\mbox{C4}}\mbox{b}$, and $\overline{\mbox{C4}}\mbox{c}$ are still semi-infinite constraints. Next, we introduce a lemma for transforming constraints $\overline{\mbox{C4}}\mbox{a}$, $\overline{\mbox{C4}}\mbox{b}$, and $\overline{\mbox{C4}}\mbox{c}$ into LMI constraints. 
\par
\textit{Lemma~1~(S-Procedure \cite{boyd2004convex}}) Let a function $f_m(\mathbf{x})$, $m\in \left \{ 1,2 \right \}$, $\mathbf{x}\in \mathbb{C}^{N\times 1}$, be defined as
\vspace*{-2mm}
\begin{equation}
f_m(\mathbf{x})= \mathbf{x}^H\mathbf{A}_m\mathbf{x}+2\Re\left \{\mathbf{a}^H_m\mathbf{x}  \right \}+a_m,\\[-2mm]
\end{equation}
where $\mathbf{A}_m\in \mathbb{H}^N$, $\mathbf{a}_m\in \mathbb{C}^{N\times 1}$, and $\mathrm{a}_m\in \mathbb{R}$. Then, the implication $f_1(\mathbf{x})\leq0 \Rightarrow f_2(\mathbf{x})\leq0$ holds if and only if there exists a $\delta \geq 0$ such that
\vspace*{-1mm}
\begin{equation}
\delta\hspace{-1mm}\begin{bmatrix}
\mathbf{A}_1 &  \mathbf{a}_1\\
\mathbf{a}_1^H &  \mathit{a}_1
\end{bmatrix}-\begin{bmatrix}
\mathbf{A}_2 &  \mathbf{a}_2\\
\mathbf{a}_2^H &  \mathit{a}_2
\end{bmatrix}\succeq \mathbf{0},\\[-1mm]
\end{equation}
provided that there exists a point $\widehat{\mathbf{x}}$ such that $f_m(\widehat{\mathbf{x}})<0$.
\par
To facilitate the application of Lemma 1, we first rewrite constraint $\overline{\mbox{C4}}\mbox{c}$ as follows
\begin{equation}
\Delta \mathbf{l}_{\mathrm{R},i}^H\mathbf{\Psi}\big(\hspace*{1mm}\underset{k\in\mathcal{K} }{\sum }\mathbf{F}\mathbf{W}_k\mathbf{F}^H+\underset{j\in\mathcal{J} }{\sum }p_j\mathbf{H}_{\mathrm{R},j}\hspace*{1mm}\big)\mathbf{\Psi}^H\Delta \mathbf{l}_{\mathrm{R},i}+\tau_i\leq
\gamma_i,\hspace*{2mm}\forall \mathbf{l}_{\mathrm{R},i}\in\Omega_{\mathrm{R},i},\hspace*{2mm}\forall i,
\end{equation}
where $\mathbf{W}_k\overset{\Delta }{=}\mathbf{w}_k\mathbf{w}_k^H$ and $\mathbf{H}_{\mathrm{R},j}\overset{\Delta }{=}\mathbf{h}_{\mathrm{R},j}\mathbf{h}_{\mathrm{R},j}^H$. By applying Lemma 1, the following implication can be obtained: $\Delta \mathbf{l}_{\mathrm{R},i}^H\Delta \mathbf{l}_{\mathrm{R},i}- \varepsilon_{\mathrm{R},i}^2\leq 0 \Rightarrow\overline{\mbox{C4}}\mbox{c}$ holds if and only if there exist $\delta_i\geq0$ such that
\begin{equation}
\hspace*{-2mm} 
\label{c4c}
\widehat{\mbox{C4}}\mbox{c:~}\mathbf{S}_{\widehat{\mathrm{C}4}\mathrm{c}_{i}}(\mathbf{W_\mathit{k}},p_j,\bm{\Psi},\gamma_i,\tau_i,\delta_i)=\begin{bmatrix}\delta_i\mathbf{I}_M-\mathbf{\Psi}\mathbf{B}\mathbf{\Psi}^H &\mathbf{0}\\
\mathbf{0} &\hspace*{2mm} -\delta_i\varepsilon_{\mathrm{R},i}^2-\tau_i+\gamma_i
\end{bmatrix}\succeq\mathbf{0},\hspace*{1mm}\forall i,
\end{equation}
where $\mathbf{B}\overset{\Delta }{=}\underset{k\in\mathcal{K} }{\sum }\mathbf{F}\mathbf{W}_k\mathbf{F}^H+\underset{j\in\mathcal{J} }{\sum }p_j\mathbf{H}_{\mathrm{R},j}$.
To simplify the notation, we rewrite the LMI in \eqref{c4c} as follows
\begin{equation}
\widehat{\mbox{C4}}\mbox{c:~}\mathbf{S}_{\widehat{\mathrm{C}4}\mathrm{c}_{i}}(\mathbf{W_\mathit{k}},p_j,\bm{\Psi},\gamma_i,\tau_i,\delta_i)=\begin{bmatrix}
\delta_i\mathbf{I}_M & \mathbf{0}\\ 
\mathbf{0} & -\delta_i\varepsilon_{\mathrm{R},i}^2-\tau_i+\gamma_i\end{bmatrix}-\mathbf{C}^H\mathbf{\Psi}\mathbf{B}\mathbf{\Psi}^H\mathbf{C}\succeq\mathbf{0},~\forall i,
\end{equation}
where $\mathbf{C}=\big [~ \mathbf{I}_M~\mathbf{0}~\big ]$. Similarly, by applying Lemma 1, we rewrite constraints $\overline{\mbox{C4}}\mbox{a}$ and $\overline{\mbox{C4}}\mbox{b}$ as follows
\begin{eqnarray}
&&\hspace*{-6mm}\widehat{\mbox{C4}}\mbox{a:~}\mathbf{S}_{\widehat{\mathrm{C}4}\mathrm{a}_{i}}(p_j,\beta_i,\iota_i)=\begin{bmatrix}\iota_i\mathbf{I}_J & \mathbf{0}\\ 
\mathbf{0} & -\iota_i\epsilon_{i,j}^2-\beta_i+\frac{p_{\mathrm{tol}_i}}{3}\end{bmatrix}-\mathbf{D}_i^H\mathbf{P}\mathbf{D}_i\succeq\mathbf{0},~\forall i,\\[2mm]
&&\hspace*{-6mm}\widehat{\mbox{C4}}\mbox{b:~}\mathbf{S}_{\widehat{\mathrm{C}4}\mathrm{b}_{i}}(\mathbf{W_\mathit{k}},\beta_i,\gamma_i,\kappa_i)=\begin{bmatrix}\kappa_i\mathbf{I}_{N_{\mathrm{T}}} &\mathbf{0}\\ 
\mathbf{0} & -\kappa_i\varepsilon_{\mathrm{D},i}^2-\gamma_i+\beta_i\end{bmatrix}-\underset{k\in\mathcal{K} }{\sum }\mathbf{E}_i^H\mathbf{W}_k\mathbf{E}_i\succeq\mathbf{0},~\forall i,
\end{eqnarray}
where $\iota_i$, $\kappa_i\geq0$, $\mathbf{P}\overset{\Delta }{=}\mathrm{diag}(p_1,\cdots,p_J)$, $\mathbf{D}_i\overset{\Delta }{=}\big [~ \mathbf{I}_J~\mathbf{0}~\big ]$, and $\mathbf{E}_i\overset{\Delta }{=}\big [~ \mathbf{I}_{N_{\mathrm{T}}}~\mathbf{0}~\big ]$.
We note that $\widehat{\mbox{C4}}\mbox{a}$ is convex with respect to $p_j$ and $\widehat{\mbox{C4}}\mbox{b}$ is convex with respect to $\mathbf{W}_k$. Moreover, $\widehat{\mbox{C4}}\mbox{c}$ is convex with respect to $p_j$ and $\mathbf{W}_k$ but is still non-convex with respect to $\bm{\Psi}$ due to the quadratic term $\mathbf{\Psi}\mathbf{B}\mathbf{\Psi}^H$. 
\subsection{Optimizing $\left \{ \mathbf{W}_k, p_j \right \}$ for Given $\bm{\Psi}$ and $\mathbf{v}_j$}
To facilitate resource allocation algorithm design, for given $\bm{\Psi}$ and $\mathbf{v}_j$, we first rewrite the terms $\left | \mathbf{g}_{\mathrm{D},k}^H\mathbf{w}_r+\mathbf{g}_{\mathrm{R},k}^H \mathbf{\Psi}\mathbf{F}\mathbf{w}_r\right |^2$ and $\left | q_{j,k}+\mathbf{g}_{\mathrm{R},k}^H \mathbf{\Psi}\mathbf{h}_{\mathrm{R},j}\right |^2$ in \eqref{downlinkR}, the term $\left | \mathbf{h}_{\mathrm{D},t}^H\mathbf{v}_j+\mathbf{h}_{\mathrm{R},t}^H \mathbf{\Psi}^H\mathbf{F}\mathbf{v}_j\right |^2$ in \eqref{uplinkR}, and the terms $\left | \overline{\mathbf{l}}_{\mathrm{D},i}^H\mathbf{w}_k +\overline{\mathbf{l}}_{\mathrm{R},i}^H\mathbf{\Psi }\mathbf{F}\mathbf{w}_k \right |^2$ and $\left | \overline{e}_{i,j}+ \overline{\mathbf{l}}_{\mathrm{R},i}^H\mathbf{\Psi }\mathbf{h}_{\mathrm{R},j} \right |^2$ in \eqref{c4d} as follows, respectively,
\vspace*{-1mm}
\begin{eqnarray}
&&\hspace*{-6mm}\left | \mathbf{g}_{\mathrm{D},k}^H\mathbf{w}_k+\mathbf{g}_{\mathrm{R},k}^H \mathbf{\Psi}\mathbf{F}\mathbf{w}_k\right |^2=\left |\widehat{\mathbf{g}}_k^H\mathbf{w}_k \right |^2=\mathrm{Tr}(\widehat{\mathbf{g}}_k\widehat{\mathbf{g}}_k^H\mathbf{W}_k),\\
&&\hspace*{-6mm}\left | \mathbf{h}_{\mathrm{D},j}^H\mathbf{v}_j+\mathbf{h}_{\mathrm{R},j}^H \mathbf{\Psi}^H\mathbf{F}\mathbf{v}_j\right |^2=\left |\widehat{\mathbf{h}}_j^H\mathbf{v}_j \right |^2=\mathrm{Tr}(\widehat{\mathbf{h}}_j\widehat{\mathbf{h}}_j^H\mathbf{v}_j\mathbf{v}_j^H),\label{wideth}\\
&&\hspace*{-6mm}\left | \overline{\mathbf{l}}_{\mathrm{D},i}^H\mathbf{w}_k +\overline{\mathbf{l}}_{\mathrm{R},i}^H\mathbf{\Psi }\mathbf{F}\mathbf{w}_k \right |^2=\left |\widehat{\mathbf{l}}_i^H\mathbf{w}_k \right |^2=\mathrm{Tr}(\widehat{\mathbf{l}}_i\widehat{\mathbf{l}}_i^H\mathbf{W}_k),\\
&&\hspace*{-6mm}\left | q_{j,k}+\mathbf{g}_{\mathrm{R},k}^H \mathbf{\Psi}\mathbf{h}_{\mathrm{R},j}\right |^2=\left |\varphi_{j,k}\right |^2,\\
&&\hspace*{-6mm}\left | \overline{e}_{i,j}+ \overline{\mathbf{l}}_{\mathrm{R},i}^H\mathbf{\Psi }\mathbf{h}_{\mathrm{R},j} \right |^2=\left |\vartheta_{i,j}\right |^2,
\end{eqnarray}
where $\widehat{\mathbf{g}}_k\in \mathbb{C}^{N_{\mathrm{T}}\times 1}$, $\widehat{\mathbf{h}}_j\in \mathbb{C}^{N_{\mathrm{T}}\times 1}$, $\widehat{\mathbf{l}}_i\in \mathbb{C}^{N_{\mathrm{T}}\times 1}$, $\varphi_{j,k}\in \mathbb{C}$, and $\vartheta_{i,j}\in \mathbb{C}$ are defined as $\widehat{\mathbf{g}}_k=\mathbf{g}_{\mathrm{D},k}+ \mathbf{F}^H\mathbf{\Psi}^H\mathbf{g}_{\mathrm{R},k}$, $\widehat{\mathbf{h}}_j=\mathbf{h}_{\mathrm{D},j}+ \mathbf{F}^H\mathbf{\Psi}\mathbf{h}_{\mathrm{R},j}$, $\widehat{\mathbf{l}}_i=\overline{\mathbf{l}}_{\mathrm{D},i}+\mathbf{F}^H\mathbf{\Psi }^H\overline{\mathbf{l}}_{\mathrm{R},i}$, $\varphi_{j,k}=q_{j,k}+\mathbf{g}_{\mathrm{R},j}^H \mathbf{\Psi}\mathbf{h}_{\mathrm{R},j}$, and $\vartheta_{i,j}=\overline{e}_{i,j}+ \overline{\mathbf{l}}_{\mathrm{R},i}^H\mathbf{\Psi }\mathbf{h}_{\mathrm{R},j}$, respectively. Then, the received SINR of the $k$-th secondary DL user and the received SINR of the $j$-th secondary UL user can be expressed as follows
\begin{eqnarray}
&&\hspace*{-12mm}\Gamma_k^{\mathrm{DL}}=\frac{\mathrm{Tr}(\widehat{\mathbf{g}}_k\widehat{\mathbf{g}}_k^H\mathbf{W}_k)}{\underset{r\in\mathcal{K}\setminus \left \{ k \right \} }{\sum} \mathrm{Tr}(\widehat{\mathbf{g}}_k\widehat{\mathbf{g}}_k^H\mathbf{W}_r)+\underset{j\in\mathcal{J}}{\sum}p_j \left | \varphi_{j,k}\right |^2+\sigma^2_{\mathrm{n}_k}}~\mbox{and}\\
&&\hspace*{-12mm}\Gamma_j^{\mathrm{UL}}=\frac{p_j\mathrm{Tr}(\widehat{\mathbf{h}}_j\widehat{\mathbf{h}}_j^H\mathbf{v}_j\mathbf{v}_j^H)}{\underset{t\in\mathcal{J}\setminus \left \{ j \right \} }{\sum} p_t\mathrm{Tr}(\widehat{\mathbf{h}}_t\widehat{\mathbf{h}}_t^H\mathbf{v}_j\mathbf{v}_j^H)+\mathrm{Tr}\big (\eta\mathbf{v}_j\mathbf{v}_j^H\mathrm{Diag}\big (\underset{k\in\mathcal{K}}{\sum }\mathbf{S}\mathbf{W}_k\mathbf{S}^H\big )\big )+\sigma^2_{\mathrm{U}}\left \| \mathbf{v}_j \right \|^2},
\end{eqnarray}
respectively. Constraint $\overline{\mbox{C4}}\mbox{d}$ can be rewritten equivalently as:
\begin{equation}
\label{c4a1}
\widehat{\mbox{C4}}\mbox{d:~}\underset{k\in \mathcal{K} }{\sum }\mathrm{Tr}(\widehat{\mathbf{l}}_i\widehat{\mathbf{l}}_i^H\mathbf{W}_k)+\underset{j\in \mathcal{J} }{\sum }p_j\left |\vartheta_{i,j}\right |^2\leq\tau_i,~\forall i.
\end{equation}
\par
Then, the joint DL transmit beamforming and UL power allocation design, i.e., $\left \{\mathbf{W}_k,p_j\right \}$, is formulated as follows
\begin{eqnarray}
\label{prob2}
&&\hspace*{-2mm}\underset{\substack{\mathbf{W}_k\in\mathbb{H}^{N_{\mathrm{T}}},p_j,\beta_i,\\\gamma_i,\tau_i,\delta_i,\iota_i,\kappa_i}}{\maxo} \,\, \,\, \underset{ j\in\mathcal{J}}{\sum}\omega_j^{\mathrm{UL}}\mathrm{log}_2(1+\Gamma_j^{\mathrm{UL}})+\underset{ k\in\mathcal{K}}{\sum}\omega_k^{\mathrm{DL}}\mathrm{log}_2(1+\Gamma_k^{\mathrm{DL}})\\
\mbox{s.t.}\hspace*{-2mm}
&&\mbox{C1:~}\underset{k\in\mathcal{K}}{\sum }\mathrm{Tr}(\mathbf{W}_k)\leq P_{\mathrm{max}}^{\mathrm{DL}},~\mbox{C2},\widehat{\mbox{C4}}\mbox{a},\widehat{\mbox{C4}}\mbox{b},\widehat{\mbox{C4}}\mbox{c},\widehat{\mbox{C4}}\mbox{d},\notag\\
&&\mbox{C5:~}\mathbf{W}_k\succeq\mathbf{0},~\forall k,~\mbox{C6:~}\mathrm{Rank}(\mathbf{W}_k)\leq1,~\forall k.\notag
\end{eqnarray}
Here, constraints C5, C6, and $\mathbf{W}_k\in\mathbb{H}^{\mathit{N}_{\mathrm{T}}}$ are imposed to ensure that $\mathbf{W}_k=\mathbf{w}_k\mathbf{w}_k^H$ holds after optimization. The non-convexity of \eqref{prob2} originates from the objective function and the rank constraint C6. Next, we aim to obtain a suboptimal solution of \eqref{prob2} iteratively by applying SCA. For notational simplicity, we define $f_1$, $f_2$, $g_1$, and $g_2$ which are given by, respectively,
\begin{eqnarray}
\hspace{-4mm}f_1\hspace{-6mm}&&=-\underset{k\in\mathcal{K} }{\sum }\omega_k^{\mathrm{DL}}\mathrm{log}_2\left(\underset{r\in\mathcal{K} }{\sum }\mathrm{Tr}(\widehat{\mathbf{g}}_k\widehat{\mathbf{g}}_k^H\mathbf{W}_r)+\underset{j\in\mathcal{J}}{\sum}p_j \left | \varphi_{j,k}\right |^2+\sigma^2_{\mathrm{n}_k}\right),\\
\hspace{-4mm}f_2\hspace{-6mm}&&=-\underset{j\in\mathcal{J} }{\sum }\omega_j^{\mathrm{UL}}\mathrm{log}_2\left(\underset{t\in\mathcal{J} }{\sum }p_t\mathrm{Tr}(\widehat{\mathbf{h}}_t\widehat{\mathbf{h}}_t^H\mathbf{v}_j\mathbf{v}_j^H)+\sigma^2_{\mathrm{U}}\left \| \mathbf{v}_j \right \|^2+\mathrm{Tr}\big (\eta\mathbf{v}_j\mathbf{v}_j^H\mathrm{Diag}\big (\underset{k\in\mathcal{K}}{\sum }\mathbf{S}\mathbf{W}_k\mathbf{S}^H\big )\big )\right),\\
\hspace{-4mm}g_1\hspace{-6mm}&&=-\underset{k\in\mathcal{K} }{\sum }\omega_k^{\mathrm{DL}}\mathrm{log}_2\left(\underset{r\in\mathcal{K}\setminus \left \{ k \right \} }{\sum}\mathrm{Tr}(\widehat{\mathbf{g}}_k\widehat{\mathbf{g}}_k^H\mathbf{W}_r)+\underset{j\in\mathcal{J}}{\sum}p_j \left | \varphi_{j,k}\right |^2+\sigma^2_{\mathrm{n}_k}\right),\\
\hspace{-4mm}g_2\hspace{-6mm}&&=-\underset{j\in\mathcal{J} }{\sum }\omega_j^{\mathrm{UL}}\mathrm{log}_2\left(\underset{t\in\mathcal{J}\setminus \left \{ j \right \} }{\sum}p_t\mathrm{Tr}(\widehat{\mathbf{h}}_t\widehat{\mathbf{h}}_t^H\mathbf{v}_j\mathbf{v}_j^H)+\sigma^2_{\mathrm{U}}\left \| \mathbf{v}_j \right \|^2+\mathrm{Tr}\big (\eta\mathbf{v}_j\mathbf{v}_j^H\mathrm{Diag}\big (\underset{k\in\mathcal{K}}{\sum }\mathbf{S}\mathbf{W}_k\mathbf{S}^H\big )\big )\right).
\end{eqnarray}
Note that the negative objective function in \eqref{prob2} can be expressed as $f_1+f_2-g_1-g_2$.
\par
Then, in the $n$-th iteration of the SCA, for a given feasible\footnote{The superscript $n$ denotes the SCA iteration index.} point $(\mathbf{W}_k^n,p_j^n)$, we construct a global underestimator of $g_1(\mathbf{W}_k,p_j)$ as follows
\begin{eqnarray}
\label{G1s}
g_1(\mathbf{W}_k,p_j)&&\hspace*{-6mm}\geq g_1(\mathbf{W}_k^n,p_j^n)+\underset{k\in\mathcal{K}}{\sum}\mathrm{Tr}\Big(\big(\nabla_{\mathbf{W}_k}g_1(\mathbf{W}_k^n,p_j^n)\big)^H(\mathbf{W}_k-\mathbf{W}_k^n)\Big)\notag\\
&&\hspace*{-6mm}+\underset{j\in\mathcal{J}}{\sum}\nabla_{p_j}g_1(\mathbf{W}_k^n,p_j^n)(p_j-p_j^n)\overset{\Delta }{=}\widehat{g}_1(\mathbf{W}_k,p_j,\mathbf{W}_k^n,p_j^n),
\end{eqnarray}
where
\begin{equation}
\nabla_{\mathbf{W}_k}g_1=-\frac{\omega_k^{\mathrm{DL}}}{\mathrm{ln2}}\underset{t\in\mathcal{K}\setminus \left \{ k \right \} }{\sum}\frac{\widehat{\mathbf{g}}_k\widehat{\mathbf{g}}_k^H}{\underset{r\in\mathcal{K}\setminus \left \{ t \right \} }{\sum}\mathrm{Tr}(\widehat{\mathbf{g}}_k\widehat{\mathbf{g}}_k^H\mathbf{W}_r)+\underset{j\in\mathcal{J}}{\sum}p_j \left | \varphi_{j,k}\right |^2+\sigma^2_{\mathrm{n}_k}},
\end{equation}
and
\begin{equation}
\nabla_{p_j}g_1=-\frac{1}{\mathrm{ln2}}\underset{k\in\mathcal{K}}{\sum}\frac{\omega_k^{\mathrm{DL}}\left | \varphi_{j,k}\right |^2}{\underset{r\in\mathcal{K}\setminus \left \{ k \right \} }{\sum}\mathrm{Tr}(\widehat{\mathbf{g}}_k\widehat{\mathbf{g}}_k^H\mathbf{W}_r)+\underset{j\in\mathcal{J}}{\sum}p_j \left | \varphi_{j,k}\right |^2+\sigma^2_{\mathrm{n}_k}}.
\end{equation}
Similarly, for a given feasible point $(\mathbf{W}_k^n,p_j^n)$, the global underestimator of $g_2(\mathbf{W}_k,p_j)$ is given by
\begin{eqnarray}
\label{G2s}
g_2(\mathbf{W}_k,p_j)&&\hspace*{-6mm}\geq g_2(\mathbf{W}_k^n,p_j^n)+\underset{k\in\mathcal{K}}{\sum}\mathrm{Tr}\Big(\big(\nabla_{\mathbf{W}_k}g_2(\mathbf{W}_k^n,p_j^n)\big)^H(\mathbf{W}_k-\mathbf{W}_k^n)\Big)\notag\\
&&\hspace*{-6mm}+\underset{j\in\mathcal{J}}{\sum}\nabla_{p_j}g_2(\mathbf{W}_k^n,p_j^n)(p_j-p_j^n)\overset{\Delta }{=}\widehat{g}_2(\mathbf{W}_k,p_j,\mathbf{W}_k^n,p_j^n),
\end{eqnarray} 
where
\begin{equation}
\nabla_{\mathbf{W}_k}g_2=-\frac{1}{\mathrm{ln2}}\underset{j\in\mathcal{J}}{\sum}\frac{\omega_j^{\mathrm{UL}}\eta\mathbf{v}_j\mathbf{v}_j^H\mathrm{Diag}(\mathbf{S}\mathbf{S}^H)}{\underset{t\in\mathcal{J}\setminus \left \{ j \right \} }{\sum}p_t\mathrm{Tr}(\widehat{\mathbf{h}}_t\widehat{\mathbf{h}}_t^H\mathbf{v}_j\mathbf{v}_j^H)+\mathrm{Tr}\big (\eta\mathbf{v}_j\mathbf{v}_j^H\mathrm{Diag}\big (\underset{k\in\mathcal{K}}{\sum }\mathbf{S}\mathbf{W}_k\mathbf{S}^H\big )\big )+\sigma^2_{\mathrm{U}}\left \| \mathbf{v}_j \right \|^2},
\end{equation}
and
\begin{equation}
\nabla_{p_j}g_2=-\frac{\omega_j^{\mathrm{UL}}}{\mathrm{ln2}}\underset{r\in\mathcal{J}\setminus \left \{ j \right \}}{\sum}\frac{\mathrm{Tr}(\widehat{\mathbf{h}}_r\widehat{\mathbf{h}}_r^H\mathbf{v}_j\mathbf{v}_j^H)}{\underset{t\in\mathcal{J}\setminus \left \{ r \right \} }{\sum}p_t\mathrm{Tr}(\widehat{\mathbf{h}}_t\widehat{\mathbf{h}}_t^H\mathbf{v}_r\mathbf{v}_r^H)+\mathrm{Tr}\big (\eta\mathbf{v}_j\mathbf{v}_j^H\mathrm{Diag}\big (\underset{k\in\mathcal{K}}{\sum }\mathbf{S}\mathbf{W}_k\mathbf{S}^H\big )\big )+\sigma^2_{\mathrm{U}}\left \| \mathbf{v}_r \right \|^2}.
\end{equation}
\par
Then, for a given feasible point $(\mathbf{W}_k^n,p_j^n)$ in the $n$-th iteration, a lower bound of the maximization problem in \eqref{prob2} can be obtained by solving the following optimization problem
\begin{eqnarray}
\label{prob3}
&&\hspace*{-22mm}\underset{\substack{\mathbf{W}_k,p_j,\beta_i,\gamma_i,\\\tau_i,\delta_i,\iota_i,\kappa_i}}{\mino} \,\, \,\, \widehat{F}(\mathbf{W}_k,p_j)\overset{\Delta }{=}f_1+f_2-\widehat{g}_1(\mathbf{W}_k,p_j,\mathbf{W}_k^n,p_j^n)-\widehat{g}_2(\mathbf{W}_k,p_j,\mathbf{W}_k^n,p_j^n)\\
\mbox{s.t.}\hspace*{-2mm}
&&\mbox{C1},\mbox{C2},\widehat{\mbox{C4}}\mbox{a},\widehat{\mbox{C4}}\mbox{b},\widehat{\mbox{C4}}\mbox{c},\widehat{\mbox{C4}}\mbox{d},\mbox{C5},\mbox{C6}.\notag
\end{eqnarray}
We note that the remaining non-convexity of problem \eqref{prob3} stems from rank-one constraint C6. Hence, we adopt SDR and remove constraint C6. The relaxed version of problem \eqref{prob3} can now be optimally solved by standard convex solvers such as CVX \cite{grant2008cvx}. Next, we verify the tightness of SDR in the following theorem.
\par
\textit{Theorem 1:~}If $P_{\mathrm{max}}^{\mathrm{DL}}>0$, an optimal beamforming matrix $\mathbf{W}_k$ satisfying $\mathrm{Rank}(\mathbf{W}_k)\leq 1$ can always be obtained.
\par
\textit{Proof:~}Please refer to Appendix A. \qed
\par
Then, we tighten the upper bound of \eqref{prob2} by solving \eqref{prob3} iteratively. The SCA algorithm for obtaining the optimal $\mathbf{w}_k^{\dagger}$ and $p_j^{\dagger}$ of \eqref{prob2} is summarized in \textbf{Algorithm 1}. We note that \textbf{Algorithm 1} is guaranteed to converge to a locally optimal solution of \eqref{prob2} \cite{dinh2010local}.
\begin{algorithm} [t]
\caption{Successive Convex Approximation Algorithm for Obtaining $\mathbf{W}_k^{\dagger}$ and $p_j^{\dagger}$}
\begin{algorithmic}[1]
\small
\STATE Set initial point $\mathbf{W}_k^1$ and $p_j^1$, iteration index $n=1$, and error tolerance $0\leq\varepsilon_{\mathrm{SCA}}\ll1$.
\REPEAT
\STATE Solve \eqref{prob3} for given $\mathbf{W}_k^n$ and $p_j^n$ and store the intermediate solution $\mathbf{W}_k$ and $p_j$
\STATE Set $n=n+1$, $\mathbf{W}_k^n=\mathbf{W}_k$, and $p_j^n=p_j$
\UNTIL $\frac{\left | \widehat{F}(\mathbf{W}_k^n,p_j^n)-\widehat{F}(\mathbf{W}_k^{n-1},p_j^{n-1}) \right |}{\left | \widehat{F}(\mathbf{W}_k^n,p_j^n)\right |}\leq \varepsilon_{\mathrm{SCA}}$
\STATE $\mathbf{W}_k^{\dagger}=\mathbf{W}_k^n$ and $p_j^{\dagger}=p_j^n$
\end{algorithmic}
\end{algorithm}
\subsection{Optimizing $\mathbf{v}_j$ for Given $\bm{\Psi}$, $\mathbf{W}_k$, and $p_j$}
For given $\bm{\Psi}$, $\mathbf{W}_k$, and $p_j$, the UL sum rate is maximized if for each uplink user $j$, the receive beamforming vector $\mathbf{v}_j$ maximizes the corresponding receive SINR $\Gamma_j^{\mathrm{UL}}$. In particular, we can obtain the optimal receive beamforming vector $\mathbf{v}_j$ by solving the following optimization problem:
\begin{eqnarray}
\label{prob4}
&&\hspace*{-2mm}\underset{\mathbf{v}_j}{\maxo} \,\, \,\, \frac{p_j\mathbf{v}_j^H\widehat{\mathbf{h}}_j\widehat{\mathbf{h}}_j^H\mathbf{v}_j}{\mathbf{v}_j^H\mathbf{R}_{\mathrm{I}_j}\mathbf{v}_j},
\end{eqnarray}
where $\widehat{\mathbf{h}}_j$ was defined in \eqref{wideth} and $\mathbf{R}_{\mathrm{I}_j}\in\mathbb{C}^{N_{\mathrm{T}}\times N_{\mathrm{T}}}$ is defined as follows
\begin{eqnarray}
\mathbf{R}_{\mathrm{I}_j}\hspace*{-6mm}&&=\underset{t\in\mathcal{J}\setminus \left \{ j \right \} }{\sum}p_t\big(\mathbf{h}_{\mathrm{D},t}\mathbf{h}_{\mathrm{D},t}^H+\mathbf{F}^H\mathbf{\Psi}\mathbf{h}_{\mathrm{R},t}\mathbf{h}_{\mathrm{D},t}^H+\mathbf{h}_{\mathrm{D},t}\mathbf{h}_{\mathrm{R},t}^H\mathbf{\Psi}^H\mathbf{F}+\mathbf{F}^H\mathbf{\Psi}\mathbf{h}_{\mathrm{R},t}\mathbf{h}_{\mathrm{R},t}^H\mathbf{\Psi}^H\mathbf{F}\big)\notag\\
&&+\eta\mathrm{Diag}\big(\underset{k\in\mathcal{K}}{\sum}\mathbf{S}\mathbf{W}_k\mathbf{S}^H\big)+\sigma_{\mathrm{U}}^2\mathbf{I}_{N_{\mathrm{T}}}.
\end{eqnarray}
Moreover, the optimization problem in \eqref{prob4} can be recast as the following equivalent convex optimization problem \cite{gershman2010convex}
\begin{eqnarray}
\label{prob5}
&&\hspace*{-6mm}\underset{\mathbf{v}_j}{\mino} \,\, \,\, \hspace*{4mm}\mathbf{v}_j^H\mathbf{R}_{\mathrm{I}_j}\mathbf{v}_j\\
&&\mbox{s.t.}\hspace*{4mm}\mbox{C7:~}\sqrt{p_j}\mathbf{v}_j^H\widehat{\mathbf{h}}_j=1.\notag
\end{eqnarray}
The optimal solution of \eqref{prob5} is given by \cite{gershman2010convex}
\begin{equation}
\label{optimalz}
    \mathbf{v}_j^{\dagger}=\varrho_j\sqrt{p_j}\mathbf{R}_{\mathrm{I}_j}^{-1}\widehat{\mathbf{h}}_j,
\end{equation}
where $\varrho_j$ is a scalar to adjust $\mathbf{v}_j^{\dagger}$ such that equality constraint C7 is satisfied. We note that for the original problem in \eqref{prob4}, $\varrho_j$ can be omitted as it has no effect on the value of the objective function.
\subsection{Optimizing $\bm{\Psi}$ for Given $\mathbf{W}_k$, $p_j$, and $\mathbf{v}_j$}
For given $\mathbf{W}_k$, $p_j$, and $\mathbf{v}_j$, the optimization problem for the IRS phase shift design is given by
\vspace*{-1mm}
\begin{eqnarray}
\label{prob6}
&&\hspace*{-16mm}\underset{\substack{\mathbf{\Psi},\beta_i,\gamma_i,\tau_i,\\\delta_i,\iota_i,\kappa_i}}{\maxo} \,\, \,\, \underset{ j\in\mathcal{J}}{\sum}\omega_j^{\mathrm{UL}}\mathrm{log}_2(1+\Gamma_j^{\mathrm{UL}})+\underset{ k\in\mathcal{K}}{\sum}\omega_k^{\mathrm{DL}}\mathrm{log}_2(1+\Gamma_k^{\mathrm{DL}})\\
\mbox{s.t.}\hspace*{-2mm}
&&\mbox{C3},\widehat{\mbox{C4}}\mbox{a},\widehat{\mbox{C4}}\mbox{b},\widehat{\mbox{C4}}\mbox{c},\widehat{\mbox{C4}}\mbox{d}\notag.
\end{eqnarray}
We note that both the objective function and constraints C3 and $\widehat{\mbox{C4}}\mbox{c}$ are non-convex functions which makes the IRS design very challenging. Next, we first tackle the non-convex objective function in \eqref{prob6}. In particular, we rewrite the quadratic term $\left | \mathbf{g}_{\mathrm{D},k}^H\mathbf{w}_r+\mathbf{g}_{\mathrm{R},k}^H \mathbf{\Psi}\mathbf{F}\mathbf{w}_r\right |^2$ in \eqref{downlinkR} as follows:
\begin{eqnarray}
&&\left | \mathbf{g}_{\mathrm{D},k}^H\mathbf{w}_r+\mathbf{g}_{\mathrm{R},k}^H \mathbf{\Psi}\mathbf{F}\mathbf{w}_r\right |^2\notag\\
&&\hspace*{-6mm}=\mathbf{g}_{\mathrm{D},k}^H\mathbf{W}_r\mathbf{g}_{\mathrm{D},k}+2\Re\left \{  \mathbf{g}_{\mathrm{D},k}^H\mathbf{W}_r\mathbf{F}^H\mathbf{\Psi}^H\mathbf{g}_{\mathrm{R},k}\right \}+\mathbf{g}_{\mathrm{R},k}^H\mathbf{\Psi}\mathbf{F}\mathbf{W}_r\mathbf{F}^H\mathbf{\Psi}^H\mathbf{g}_{\mathrm{R},k}\notag
\\&&\hspace*{-6mm}=\mathbf{g}_{\mathrm{D},k}^H\mathbf{W}_r\mathbf{g}_{\mathrm{D},k}+2\Re\left \{ \mathbf{g}_{\mathrm{D},k}^H\mathbf{W}_r\mathbf{F}^H\mathrm{diag}(\mathbf{g}_{\mathrm{R},k})\bm{\theta}^H\right \}+\bm{\theta}^H\mathrm{diag}(\mathbf{g}_{\mathrm{R},k}^H)\mathbf{F}\mathbf{W}_r\mathbf{F}^H\mathrm{diag}(\mathbf{g}_{\mathrm{R},k})\bm{\theta}\notag
\\&&\hspace*{-6mm}=\mathrm{Tr}\Big(\begin{bmatrix}
\bm{\theta}^H & \rho^*
\end{bmatrix}\begin{bmatrix}
\mathrm{diag}(\mathbf{g}_{\mathrm{R},k}^H)\mathbf{F}\\ \mathbf{g}^H_{\mathrm{D},k}
\end{bmatrix}\mathbf{W}_r\begin{bmatrix}
\mathbf{F}^H\mathrm{diag}(\mathbf{g}_{\mathrm{R},k})& \mathbf{g}_{\mathrm{D},k}
\end{bmatrix}
\begin{bmatrix}
\bm{\theta}\\ \rho
\end{bmatrix}\Big)\notag
\\&&\hspace*{-6mm}=\mathrm{Tr}(\widetilde{\bm{\theta}}^H\mathbf{G}_{k}\mathbf{W}_r\mathbf{G}_{k}^H\widetilde{\bm{\theta}})=\mathrm{Tr}(\bm{\Theta}\mathbf{G}_{k}\mathbf{W}_r\mathbf{G}_{k}^H),
\end{eqnarray}
where optimization variables $\bm{\theta}\in\mathbb{C}^{M\times 1}$, $\widetilde{\bm{\theta}}\in\mathbb{C}^{(M+1)\times 1}$, and $\bm{\Theta}\in\mathbb{C}^{(M+1)\times(M+1)}$ are defined as $\bm{\theta}=[e^{j\psi_1}, \cdots, e^{j\psi_M}]^H$, $\widetilde{\bm{\theta}}=[\bm{\theta}^T~ \rho]^T$, and $\bm{\Theta}=\widetilde{\bm{\theta}}\widetilde{\bm{\theta}}^H$, respectively. Moreover, $\rho\in\mathbb{C}$ is a dummy variable with $\left | \rho \right |^2=1$. Besides, $\mathbf{G}_{k}\in\mathbb{C}^{(M+1)\times N_{\mathrm{T}}}$ is defined as  $\mathbf{G}_{k}=\big[\big(\mathrm{diag}(\mathbf{g}_{\mathrm{R},k}^H)\mathbf{F}\big)^T~~\mathbf{g}_{\mathrm{D},k}^*\big]^T$.
\par
Similarly, we rewrite the term $\left | q_{j,k}+\mathbf{g}_{\mathrm{R},k}^H \mathbf{\Psi}\mathbf{h}_{\mathrm{R},j}\right |^2$ in \eqref{downlinkR}, the term $\left | \mathbf{h}_{\mathrm{D},t}^H\mathbf{v}_j+\mathbf{h}_{\mathrm{R},t}^H \mathbf{\Psi}\mathbf{F}\mathbf{v}_j\right |^2$ in \eqref{uplinkR}, and the terms $\left | \overline{\mathbf{l}}_{\mathrm{D},i}^H\mathbf{w}_k +\overline{\mathbf{l}}_{\mathrm{R},i}^H\mathbf{\Psi }\mathbf{F}\mathbf{w}_k \right |^2$ and $\left | \overline{e}_{i,j}+ \overline{\mathbf{l}}_{\mathrm{R},i}^H\mathbf{\Psi }\mathbf{h}_{\mathrm{R},j} \right |^2$ in \eqref{c4d} as follows, respectively,
\begin{eqnarray}
&&\left | q_{j,k}+\mathbf{g}_{\mathrm{R},k}^H \mathbf{\Psi}\mathbf{h}_{\mathrm{R},j}\right |^2=\mathrm{Tr}(\bm{\Theta}\mathbf{Q}_{j,k}),\\
&&\left | \mathbf{h}_{\mathrm{D},t}^H\mathbf{v}_j+\mathbf{h}_{\mathrm{R},t}^H \mathbf{\Psi}^H\mathbf{F}\mathbf{v}_j\right |^2
=\mathrm{Tr}(\bm{\Theta}^T\mathbf{H}_{t}\mathbf{v}_j\mathbf{v}_j^H\mathbf{H}_{t}^H),\\
&&\left | \mathbf{l}_{\mathrm{D},i}^H\mathbf{w}_k +\overline{\mathbf{l}}_{\mathrm{R},i}^H\mathbf{\Psi }\mathbf{F}\mathbf{w}_k \right |^2=\mathrm{Tr}(\bm{\Theta}\mathbf{L}_{i}\mathbf{W}_k\mathbf{L}_{i}^H),\\
&&\left | e_{i,j}+ \overline{\mathbf{l}}_{\mathrm{R},i}^H\mathbf{\Psi }\mathbf{h}_{\mathrm{R},j} \right |^2=\mathrm{Tr}(\bm{\Theta}\mathbf{P}_{i,j}),
\end{eqnarray}
where $\mathbf{H}_{t}\in\mathbb{C}^{(M+1)\times N_{\mathrm{T}}}$ and $\mathbf{L}_{i}\in\mathbb{C}^{(M+1)\times N_{\mathrm{T}}}$ are defined as $\mathbf{H}_{t}=\big[\big(\mathrm{diag}(\mathbf{h}_{\mathrm{R},t}^H)\mathbf{F}\big)^T~~\mathbf{h}_{\mathrm{D},t}^*\big]^T$ and $\mathbf{L}_{i}=\big[\big(\mathrm{diag}(\overline{\mathbf{l}}_{\mathrm{R},i}^H)\mathbf{F}\big)^T~~\mathbf{l}_{\mathrm{D},i}^*\big]^T$, respectively. Moreover, $\mathbf{Q}_{j,k}\in\mathbb{C}^{(M+1)\times (M+1)}$ and $\mathbf{P}_{i,j}\in\mathbb{C}^{(M+1)\times (M+1)}$ are defined as 
\begin{eqnarray}
&&\mathbf{Q}_{j,k}=\begin{bmatrix}
\mathrm{diag}(\mathbf{g}_{\mathrm{R},k}^H)\mathbf{H}_{\mathrm{R},j}\mathrm{diag}(\mathbf{g}_{\mathrm{R},k}) & q_{j,k}^*\mathrm{diag}(\mathbf{g}_{\mathrm{R},k}^H)\mathbf{h}_{\mathrm{R},j}\\ \mathbf{h}_{\mathrm{R},j}^H\mathrm{diag}(\mathbf{g}_{\mathrm{R},k})q_{j,k} 
 & \left | q_{j,k} \right |^2
\end{bmatrix},\\
&&\mathbf{P}_{i,j}=\begin{bmatrix}
\mathrm{diag}(\overline{\mathbf{l}}_{\mathrm{R},i}^H)\mathbf{H}_{\mathrm{R},j}\mathrm{diag}(\overline{\mathbf{l}}_{\mathrm{R},i}) & e_{i,j}^*\mathrm{diag}(\overline{\mathbf{l}}_{\mathrm{R},i}^H)\mathbf{h}_{\mathrm{R},j}\\ \mathbf{h}_{\mathrm{R},j}^H\mathrm{diag}(\overline{\mathbf{l}}_{\mathrm{R},i})e_{i,j} 
 & \left | e_{i,j} \right |^2
 \end{bmatrix},
\end{eqnarray}
respectively.
\par
Then, we rewrite constraint $\widehat{\mbox{C4}}\mbox{d}$ equivalently as
\begin{equation}
    \widetilde{\mbox{C4}}\mbox{d:~}\underset{k\in \mathcal{K} }{\sum }\mathrm{Tr}(\bm{\Theta}\mathbf{L}_{i}\mathbf{W}_k\mathbf{L}_{i}^H)+\underset{j\in \mathcal{J} }{\sum }\mathrm{Tr}(\bm{\Theta}\mathbf{P}_{i,j})\leq\tau_i,~\forall i.
\end{equation}
We note that $\widetilde{\mbox{C4}}\mbox{d}$ is a convex constraint with respect to $\bm{\Theta}$. Moreover, the receive SINR of secondary DL user $k$ and the receive SINR of secondary UL user $j$ can be equivalently respectively rewritten as follows
\begin{eqnarray}
\label{downlinkR2}
&&\hspace*{-6mm}\Gamma_k^{\mathrm{DL}}=\frac{\mathrm{Tr}(\bm{\Theta}\mathbf{G}_{k}\mathbf{W}_k\mathbf{G}_{k}^H)}{\underset{r\in\mathcal{K}\setminus \left \{ k \right \} }{\sum}\mathrm{Tr}(\bm{\Theta}\mathbf{G}_{k}\mathbf{W}_r\mathbf{G}_{k}^H)+\underset{j\in\mathcal{J}}{\sum}p_j \mathrm{Tr}(\bm{\Theta}\mathbf{Q}_{j,k}) +\sigma^2_{\mathrm{n}_k}},\\
\label{uplinkR2}
&&\hspace*{-6mm}\Gamma_j^{\mathrm{UL}}=\frac{p_j\mathrm{Tr}(\bm{\Theta}^T\mathbf{H}_{j}\mathbf{v}_j\mathbf{v}_j^H\mathbf{H}_{j}^H)}{\underset{t\in\mathcal{J}\setminus \left \{ j \right \} }{\sum} p_t\mathrm{Tr}(\bm{\Theta}^T\mathbf{H}_{t}\mathbf{v}_j\mathbf{v}_j^H\mathbf{H}_{t}^H)+\mathrm{Tr}\big (\eta\mathbf{v}_j\mathbf{v}_j^H\mathrm{Diag}\big (\underset{k\in\mathcal{K}}{\sum }\mathbf{S}\mathbf{W}_k\mathbf{S}^H\big )\big )+\sigma^2_{\mathrm{U}}\left \| \mathbf{v}_j\right \|^2}.
\end{eqnarray}
\par
The objective function of the negative optimization problem in \eqref{prob6} can be rewritten as:
\vspace*{-1mm}
\begin{eqnarray}
\underset{ j\in\mathcal{J}}{\sum}\omega_j^{\mathrm{UL}}\mathrm{log}_2(1+\Gamma_j^{\mathrm{UL}})+\underset{ k\in\mathcal{K}}{\sum}\omega_k^{\mathrm{DL}}\mathrm{log}_2(1+\Gamma_k^{\mathrm{DL}})= \widetilde{f}_1+\widetilde{f}_2-\widetilde{g}_1-\widetilde{g}_2,
\end{eqnarray}
where $f_1$, $f_2$, $g_1$, and $g_2$ are given by, respectively,
\begin{eqnarray}
\hspace*{-4mm}\widetilde{f}_1&&\hspace*{-6mm}=-\underset{k\in\mathcal{K}}{\sum }\omega_k^{\mathrm{DL}}\mathrm{log}_2\Big(\underset{r\in\mathcal{K}}{\sum}\mathrm{Tr}(\bm{\Theta}\mathbf{G}_{k}\mathbf{W}_r\mathbf{G}_{k}^H)+\underset{j\in\mathcal{J}}{\sum}p_j \big(\mathrm{Tr}(\bm{\Theta}\mathbf{Q}_{j,k})\big) +\sigma^2_{\mathrm{n}_k}\Big),\\
\hspace*{-4mm}\widetilde{f}_2&&\hspace*{-6mm}=-\underset{j\in\mathcal{J}}{\sum }\omega_j^{\mathrm{UL}}\mathrm{log}_2\Big(\underset{t\in\mathcal{J}}{\sum}p_t\mathrm{Tr}(\bm{\Theta}^T\mathbf{H}_{t}\mathbf{v}_j\mathbf{v}_j^H\mathbf{H}_{t}^H)+\mathrm{Tr}\big (\eta\mathbf{v}_j\mathbf{v}_j^H\mathrm{Diag}\big (\underset{k\in\mathcal{K}}{\sum }\mathbf{S}\mathbf{W}_k\mathbf{S}^H\big )\big )+\sigma^2_{\mathrm{U}}\left \| \mathbf{v}_j\right \|^2\Big),\\
\hspace*{-4mm}\widetilde{g}_1&&\hspace*{-6mm}=-\underset{k\in\mathcal{K}}{\sum }\omega_k^{\mathrm{DL}}\mathrm{log}_2\Big(\underset{r\in\mathcal{K}\setminus \left \{ k \right \} }{\sum}\mathrm{Tr}(\bm{\Theta}\mathbf{G}_{k}\mathbf{W}_r\mathbf{G}_{k}^H)+\underset{j\in\mathcal{J}}{\sum}p_j \big(\mathrm{Tr}(\bm{\Theta}\mathbf{Q}_{j,k})\big) +\sigma^2_{\mathrm{n}_k}\Big),\\
\hspace*{-4mm}\widetilde{g}_2&&\hspace*{-6mm}=-\underset{j\in\mathcal{J}}{\sum }\omega_j^{\mathrm{UL}}\mathrm{log}_2\Big(\underset{t\in\mathcal{J}\setminus \left \{ j \right \} }{\sum}p_t\mathrm{Tr}(\bm{\Theta}^T\mathbf{H}_{t}\mathbf{v}_j\mathbf{v}_j^H\mathbf{H}_{t}^H)\hspace{-1mm}+\mathrm{Tr}\big (\eta\mathbf{v}_j\mathbf{v}_j^H\mathrm{Diag}\big (\underset{k\in\mathcal{K}}{\sum }\mathbf{S}\mathbf{W}_k\mathbf{S}^H\big )\big )\hspace{-1mm}+\hspace{-1mm}\sigma^2_{\mathrm{U}}\left \| \mathbf{v}_j\right \|^2\Big).
\end{eqnarray}
\par
Next, by employing singular value decomposition, we transform constraint $\widehat{\mbox{C4}}\mbox{c}$ into a convex constraint. Specifically, for given $\mathbf{W}_k$ and $p_j$, we recast matrix $\mathbf{B}$ as $\mathbf{B}=\underset{s}{\sum }\widetilde{\sigma}_s\widetilde{\mathbf{u}}_s\widetilde{\mathbf{v}}_s^H$, where $\widetilde{\sigma}_s$ are the singular values of $\mathbf{B}$, and $\widetilde{\mathbf{u}}_d$ and $\widetilde{\mathbf{v}}_d$ are the corresponding left and right singular vectors of $\mathbf{B}$, respectively. Then, we rewrite the term $\mathbf{C}^H\mathbf{\Psi}\mathbf{B}\mathbf{\Psi}^H\mathbf{C}$ in constraint $\widehat{\mbox{C4}}\mbox{c}$ as follows
\vspace*{-2mm}
\begin{equation}
\mathbf{C}^H\mathbf{\Psi}\mathbf{B}\mathbf{\Psi}^H\mathbf{C}=\underset{s}{\sum}\widetilde{\sigma}_s 
\mathbf{C}^H\mathrm{diag}(\widetilde{\mathbf{u}}_s)\bm{\theta}\bm{\theta}^H\mathrm{diag}(\widetilde{\mathbf{v}}_s^H)\mathbf{C}=\underset{s}{\sum}\widetilde{\sigma}_s\widetilde{\mathbf{D}}_s\bm{\Theta}\widetilde{\mathbf{E}}_s,
\end{equation}
where $\widetilde{\mathbf{D}}_s$, $\widetilde{\mathbf{E}}_s\in\mathbb{C}^{(M+1)\times(M+1)}$ are defined as $\widetilde{\mathbf{D}}_s=\begin{bmatrix}
\mathbf{C}^H\mathrm{diag}(\widetilde{\mathbf{u}}_s) & \mathbf{0}
\end{bmatrix}$ and $\widetilde{\mathbf{E}}_s=\begin{bmatrix}
\mathrm{diag}(\widetilde{\mathbf{v}}_s)\mathbf{C} \\ \mathbf{0}
\end{bmatrix}$, respectively. Hence, we can rewrite constraint $\widehat{\mbox{C4}}\mbox{c}$ equivalently as
\vspace*{-2mm}
\begin{equation}
\widetilde{\mbox{C4}}\mbox{c:}\hspace*{1mm}\mathbf{S}_{\overline{\mathrm{C}4}\mathrm{c}_{i}}(\bm{\Theta},\tau_i,\delta_i)=\begin{bmatrix}
\delta_i\mathbf{I}_M & \mathbf{0}\\ 
\mathbf{0} & -\delta_i\varepsilon_{\mathrm{R},i}^2-\tau_i+\gamma_i\end{bmatrix}-\underset{s}{\sum}\widetilde{\sigma}_s\widetilde{\mathbf{D}}_s\bm{\Theta}\widetilde{\mathbf{E}}_s\succeq\mathbf{0},~\forall i.
\end{equation}
Now, constraint $\widetilde{\mbox{C4}}\mbox{c}$ is a convex function with respect to $\bm{\Theta}$.
\par
Therefore, for given $\mathbf{W}_k$, $p_j$, and $\mathbf{v}_j$, we can obtain $\bm{\Theta}$ by solving the following optimization problem
\vspace*{-4mm}
\begin{eqnarray}
\label{prob7}
&&\hspace*{-2mm}\underset{\substack{\mathbf{\Theta}\in\mathbb{H}^{M+1},\beta_i,\gamma_i,\\\tau_i,\delta_i,\iota_i,\kappa_i}}{\mino} \,\, \,\, \widetilde{f}_1+\widetilde{f}_2-\widetilde{g}_1-\widetilde{g}_2\\
\mbox{s.t.}\hspace*{-2mm}
&&\widetilde{\mbox{C3}}\mbox{:}~
\mathrm{Diag}(\mathbf{\Theta})=\mathbf{I}_{M+1},~\widehat{\mbox{C4}}\mbox{a},\widehat{\mbox{C4}}\mbox{b}.\widetilde{\mbox{C4}}\mbox{c},\widetilde{\mbox{C4}}\mbox{d},\notag\\
&&\mbox{C8:~}\mathbf{\Theta}\succeq\mathbf{0},~\mbox{C9:~}\mathrm{Rank}(\mathbf{\Theta})=1,\notag
\end{eqnarray}
where $\bm{\Theta}\succeq\mathbf{0}$ and constraints C8 and C9 are imposed to ensure $\bm{\Theta}=\widetilde{\bm{\theta}}\widetilde{\bm{\theta}}^H$ holds after optimization. We note that the rank-one constraint C9 is an obstacle to solving problem \eqref{prob7}. In the literature, SDR is commonly adopted to tackle the rank-one constraint \cite{5447068}. Yet, applying SDR to \eqref{prob7} may not result in a rank-one matrix $\mathbf{\Theta}$. Moreover, some approximation methods such as Gaussian randomization cannot guarantee the convergence of the overall BCD algorithm \cite{5447068}. To tackle this obstacle, we first transform the combinatorial constraint C9 equivalently into the following difference of convex (d.c.) functions constraint \cite{8761429}:
\begin{equation}
    \widetilde{\mbox{C9}}\mbox{:}~\left \|\bm{\Theta}\right \|_*-\left \| \mathbf{\Theta } \right \|_2\leq 0,
\end{equation}
where $\left \| \mathbf{\Theta } \right \|_2$ denotes the spectral norm, i.e., $\left \| \mathbf{\Theta } \right \|_2=\sigma_1(\mathbf{\Theta })$, where $\sigma_i(\mathbf{\Theta })$ denotes the $i$-th largest singular value of matrix $\mathbf{\Theta }$. We note that for any $\bm{\Theta}\in\mathbb{H}^{M+1}$ and $\bm{\Theta}\succeq\mathbf{0}$, we have $\left \|\bm{\Theta}\right \|_*=\underset{i}{\sum}~\sigma_i(\mathbf{\Theta })\geq\left \| \mathbf{\Theta } \right \|_2=\underset{i}{\mathrm{max}}~\sigma_i(\mathbf{\Theta })$ and the equality holds if and only if $\bm{\Theta}$ is a rank-one matrix. Yet, the resulting constraint $\widetilde{\mbox{C9}}$ is still non-convex. To circumvent this obstacle, we adopt a penalty approach \cite{ben1997penalty} and recast \eqref{prob7} as follows:
\begin{eqnarray}
\label{prob8}
&&\hspace*{-26mm}\underset{\substack{\mathbf{\Theta}\in\mathbb{H}^{M+1},\beta_i,\gamma_i,\\\tau_i,\delta_i,\iota_i,\kappa_i}}{\mino} \,\, \,\, \widetilde{f}_1+\widetilde{f}_2-\widetilde{g}_1-\widetilde{g}_2+\chi\big(\left \|\bm{\Theta}\right \|_*-\left \| \mathbf{\Theta } \right \|_2\big)\\
\mbox{s.t.}\hspace*{4mm}
&&\widetilde{\mbox{C3}},\widehat{\mbox{C4}}\mbox{a},\widehat{\mbox{C4}}\mbox{b},\widetilde{\mbox{C4}}\mbox{c},\widetilde{\mbox{C4}}\mbox{d},\mbox{C8},\notag
\end{eqnarray}
where $\chi\gg0$ is a constant which penalizes the objective function for any matrix $\bm{\Theta}$ whose rank is larger than one. Then, we use a sequence of $\chi_q$ to approach infinity and reveal that problem \eqref{prob8} is equivalent to problem \eqref{prob7} in the following theorem \cite{ben1997penalty}.
\par
\textit{Theorem 2:~} Denote the optimal solution of problem \eqref{prob8} as $\bm{\Theta}_q$ with penalty factor $\chi_q$. When $\chi_q$ is sufficiently large, i.e., $\chi_q\rightarrow \infty$, every limit point $\overline{\bm{\Theta}}$ of the sequence $\left \{ \bm{\Theta}_q \right \}$ is an optimal solution of problem \eqref{prob7}.
\par
\textit{Proof:~}Please refer to Appendix B. \qed
\par 
The optimization problem in \eqref{prob8} is still an intractable problem due to the non-convexity of the objective function. Yet, we note that $\widetilde{f}_1$, $\widetilde{f}_2$, $\widetilde{g}_1$, $\widetilde{g}_2$, $\left \|\bm{\Theta}\right \|_*$, and $\left \| \mathbf{\Theta } \right \|_2$ are all convex functions and the problem in \eqref{prob8} is in the canonical form of d.c. programming. Thus, a stationary point of \eqref{prob8} can be obtained by applying SCA \cite{dinh2010local}. To start with, we first construct a global underestimator of $\widetilde{g}_1$. In particular, for any feasible point $\bm{\Theta}^n$, the differentiable convex function $\widetilde{g}_1(\bm{\Theta})$ satisfies the following inequality:
\begin{eqnarray}
\label{G1}
\widetilde{g}_1(\bm{\Theta})&&\hspace*{-6mm}\geq \widetilde{g}_1(\bm{\Theta}^n)+\mathrm{Tr}\Big(\big(\nabla_{\bm{\Theta}}\widetilde{g}_1(\bm{\Theta}^n)\big)^H(\bm{\Theta}-\bm{\Theta}^n)\Big)\overset{\Delta }{=}\overline{g}_1(\bm{\Theta},\bm{\Theta}^n),\\[-7mm]\notag
\end{eqnarray}
where $\nabla_{\bm{\Theta}}\widetilde{g}_1$ is given by
\begin{equation}
   \nabla_{\bm{\Theta}}\widetilde{g}_1=-\frac{1}{\mathrm{ln2}}\underset{k\in\mathcal{K}}{\sum }\frac{\omega_k^{\mathrm{DL}}\underset{r\in\mathcal{K}\setminus \left \{ k \right \} }{\sum}\mathbf{G}_{k}\mathbf{W}_r^H\mathbf{G}_{k}^H+\underset{j\in\mathcal{J}}{\sum}p_j\mathbf{Q}^H_{j,k}}{\underset{r\in\mathcal{K}\setminus \left \{ k \right \} }{\sum}\mathrm{Tr}(\bm{\Theta}\mathbf{G}_{k}\mathbf{W}_r\mathbf{G}_{k}^H)+\underset{j\in\mathcal{J}}{\sum}p_j \big(\mathrm{Tr}(\bm{\Theta}\mathbf{Q}_{j,k})\big) +\sigma^2_{\mathrm{n}_k}}, 
\end{equation}
and $\overline{g}_1(\bm{\Theta},\bm{\Theta}^n)$ in \eqref{G1} is a global underestimator of $\widetilde{g}_1(\bm{\Theta})$. Similarly, for feasible point $\bm{\Theta}^n$, global underestimators of $\widetilde{g}_2(\bm{\Theta})$ and $\left \| \bm{\Theta} \right \|_2$ can be constructed as follows, respectively,
\begin{eqnarray}
\label{G2}
\widetilde{g}_2(\bm{\Theta})&&\hspace*{-6mm}\geq \widetilde{g}_2(\bm{\Theta}^n)+\mathrm{Tr}\Big(\big(\nabla_{\bm{\Theta}}\widetilde{g}_2(\bm{\Theta}^n)\big)^H(\bm{\Theta}-\bm{\Theta}^n)\Big)\overset{\Delta }{=}\overline{g}_2(\bm{\Theta},\bm{\Theta}^n),\\[-7mm]\notag
\end{eqnarray}
where $\nabla_{\bm{\Theta}}\widetilde{g}_2$ is given by
\begin{equation}
   \nabla_{\bm{\Theta}}\widetilde{g}_2\hspace*{-1mm}=\hspace*{-1mm}-\underset{j\in\mathcal{J}}{\sum }\frac{\frac{\omega_j^{\mathrm{UL}}}{\mathrm{ln2}}\underset{t\in\mathcal{J}\setminus \left \{ j \right \} }{\sum}p_t\mathbf{H}_{t} \mathbf{v}_j\mathbf{v}_j^H\mathbf{H}_{t}^H}{\hspace*{-2mm}\underset{t\in\mathcal{J}\setminus \left \{ j \right \} }{\sum}\hspace*{-3mm}p_t\big(\mathrm{Tr}(\bm{\Theta}^T\mathbf{H}_{t}\mathbf{v}_j\mathbf{v}_j^H\mathbf{H}_{t}^H)\big)+\mathrm{Tr}\big (\eta\mathbf{v}_j\mathbf{v}_j^H\mathrm{Diag}\big (\underset{k\in\mathcal{K}}{\sum }\mathbf{S}\mathbf{W}_k\mathbf{S}^H\big )\big )+\sigma^2_{\mathrm{U}}\left \| \mathbf{v}_j\right \|^2},
\end{equation}
and 
\begin{eqnarray}
\label{Theta}
\left \| \bm{\Theta } \right \|_2&&\hspace*{-6mm}\geq \left \| \bm{\Theta}^n \right \|_2+\mathrm{Tr}\Big(\bm{\theta}^n_{\mathrm{max}}(\bm{\theta}^n_{\mathrm{max}})^H(\bm{\Theta}-\bm{\Theta}^n)\Big)\overset{\Delta }{=}\overline{\Theta}^n,\\[-7mm]\notag
\end{eqnarray}
where $\bm{\theta}^n_{\mathrm{max}}$ is the eigenvector associated with the principal eigenvalue of $\mathbf{\Theta}^n$.
\par
\begin{algorithm}[t]
\caption{Successive Convex Approximation Algorithm for Obtaining $\mathbf{\Psi}^{\dagger}$}
\begin{algorithmic}[1]
\small
\STATE Set initial point $\mathbf{\Theta }^1$, iteration index $n=1$, and error tolerance $0\leq\varepsilon_{\mathrm{SCA}}\ll1$.
\REPEAT
\STATE For given $\mathbf{\Theta }^n$, obtain the intermediate solution $\mathbf{\Theta }$ by solving \eqref{prob9}
\STATE Set $n=n+1$ and $\mathbf{\Theta }^n=\mathbf{\Theta }$
\UNTIL $\frac{\left | \widetilde{F}(\mathbf{\Theta}^n)-\widetilde{F}(\mathbf{\Theta}^{n-1}) \right |}{\left | \widetilde{F}(\mathbf{\Theta}^n)\right |}\leq \varepsilon_{\mathrm{SCA}}$
\STATE $\mathbf{\Theta }^{\dagger}=\mathbf{\Theta }^n$
\STATE Recover $\mathbf{\Psi}^{\dagger}$ from $\mathbf{\Theta}^{\dagger}$
\end{algorithmic}
\end{algorithm}
Therefore, for any given point $\mathbf{\Theta}^n$, an upper bound on \eqref{prob7} is obtained by solving the following optimization problem:
\vspace*{-3mm}
\begin{eqnarray}
\label{prob9}
&&\hspace*{-26mm}\underset{\substack{\mathbf{\Theta}\in\mathbb{H}^{M+1},\beta_i,\gamma_i,\\\tau_i,\delta_i,\iota_i,\kappa_i}}{\mino} \,\, \,\, \widetilde{F}(\mathbf{\Theta})\overset{\Delta }{=}\widetilde{f}_1+\widetilde{f}_2-\overline{g}_1-\overline{g}_2+\chi\big(\left \|\bm{\Theta}\right \|_*-\overline{\Theta}^n\big)\\
\mbox{s.t.}\hspace*{4mm}
&&\widetilde{\mbox{C3}},\widehat{\mbox{C4}}\mbox{a},\widehat{\mbox{C4}}\mbox{b}.\widetilde{\mbox{C4}}\mbox{c},\widetilde{\mbox{C4}}\mbox{d},\mbox{C8}.\notag
\end{eqnarray}
Note that \eqref{prob9} is a convex optimization problem and the optimal solution of \eqref{prob9} can be obtained via CVX \cite{grant2008cvx}. The proposed algorithm for solving \eqref{prob8} is summarized in \textbf{Algorithm 2}. We note that the function value of \eqref{prob8} is upper bounded by the minimum of \eqref{prob9}. Moreover, by iteratively applying \textbf{Algorithm 2}, we can gradually tighten the upper bound and obtain a sequence of solutions $\bm{\Theta}$. Furthermore, the objective function of \eqref{prob9} is monotonically non-increasing and the developed algorithm is guaranteed to converge to a stationary point of \eqref{prob8} \cite{dinh2010local}.
\begin{algorithm}[t]
\caption{Block Coordinate Descent Algorithm}
\begin{algorithmic}[1]
\small
\STATE Set initial points $(\mathbf{w}_k)^1$, $(p_j)^1$, $(\mathbf{v}_j)^1$, and $(\mathbf{\Psi})^1$, iteration index $m=1$, and convergence tolerance $0\leq\varepsilon_{\mathrm{BCD}}\ll1$
\REPEAT
\STATE Solve \eqref{prob3} for given $\mathbf{\Psi}=(\mathbf{\Psi})^m$ and $\mathbf{v}_j=(\mathbf{v}_j)^m$ by applying \textbf{Algorithm 1} and obtain $(\mathbf{w}_k)^{m+1}$ and $(p_j)^{m+1}$
\STATE Calculate $(\mathbf{v}_j)^{m+1}$ for given $\mathbf{\Psi}=(\mathbf{\Psi})^m$, $\mathbf{w}_k=(\mathbf{w}_k)^{m+1}$, and $p_j=(p_j)^{m+1}$ using \eqref{optimalz}
\STATE Solve \eqref{prob9} for $\mathbf{w}_k=(\mathbf{w}_k)^{m+1}$, $p_j=(p_j)^{m+1}$, and $\mathbf{v}_j=(\mathbf{v}_j)^{m+1}$ by applying \textbf{Algorithm 2} and recover $(\mathbf{\Psi})^{m+1}$ based on $(\mathbf{\Theta})^{m+1}$
\STATE Set $m=m+1$
\UNTIL $\frac{\left | F\Big((\mathbf{w}_k)^m, (p_j)^m,(\mathbf{v}_j)^m,(\mathbf{\Psi})^m\Big)-F\Big((\mathbf{w}_k)^{m-1}, (p_j)^{m-1},(\mathbf{v}_j)^{m-1},(\mathbf{\Psi})^{m-1}\Big) \right |}{\left |F\Big((\mathbf{w}_k)^{m-1}, (p_j)^{m-1},(\mathbf{v}_j)^{m-1},(\mathbf{\Psi})^{m-1}\Big)  \right |}\leq \varepsilon_{\mathrm{BCD}}$, where $F(\cdot,\cdot,\cdot,\cdot)$ is defined in \eqref{prob1}
\STATE $\mathbf{w}_k^{\dagger}=(\mathbf{w}_k)^m$, $p_j^{\dagger}=(p_j)^m$, $\mathbf{v}_j^{\dagger}=(\mathbf{v}_j)^m$, $\mathbf{\Psi}^{\dagger}=(\mathbf{\Psi})^m$
\end{algorithmic}
\end{algorithm}
\par
The overall BCD based algorithm is summarized in \textbf{Algorithm 3}. Recall that objective function in \eqref{prob3} is monotonically decreasing in each iteration of \textbf{Algorithm 1} and the receive beamforming vector $\mathbf{v}_j^{\dagger}$ admits a closed-form solution, cf. \eqref{optimalz}. We note that any limit point of the non-increasing sequence $\left \{(\mathbf{w}_k)^m, (p_j)^m,(\mathbf{v}_j)^m,(\mathbf{\Psi})^m\right \}_{m\in\mathbb{N}}$ obtained with \textbf{Algorithm 3} is a stationary point of \eqref{altprob1}. Moreover, the function value of the sequence $\left \{(\mathbf{w}_k)^m, (p_j)^m,(\mathbf{v}_j)^m,(\mathbf{\Psi})^m\right \}_{m\in\mathbb{N}}$ is guaranteed to converge to a stationary value of the objective function of \eqref{altprob1} in polynomial time \cite{tseng2001convergence}. We note that, due to the safe approximation of constraint C4, a stationary point of \eqref{altprob1} is a feasible suboptimal solution of the original problem in \eqref{prob1}. Besides, the per iteration computational complexity of the developed BCD algorithm is given by $\mathcal{O}\Big(2I\big(N_{\mathrm{T}}^3+(M+1)^3\big)+4I^2\big(N_{\mathrm{T}}^2+(M+1)^2\big)+2I^3\Big)$ \cite[Theorem 3.12]{polik2010interior}.
\section{Simulation Results}
\begin{figure}[t]
\begin{minipage}[b]{0.47\linewidth} 
\centering\includegraphics[width=3.2in]{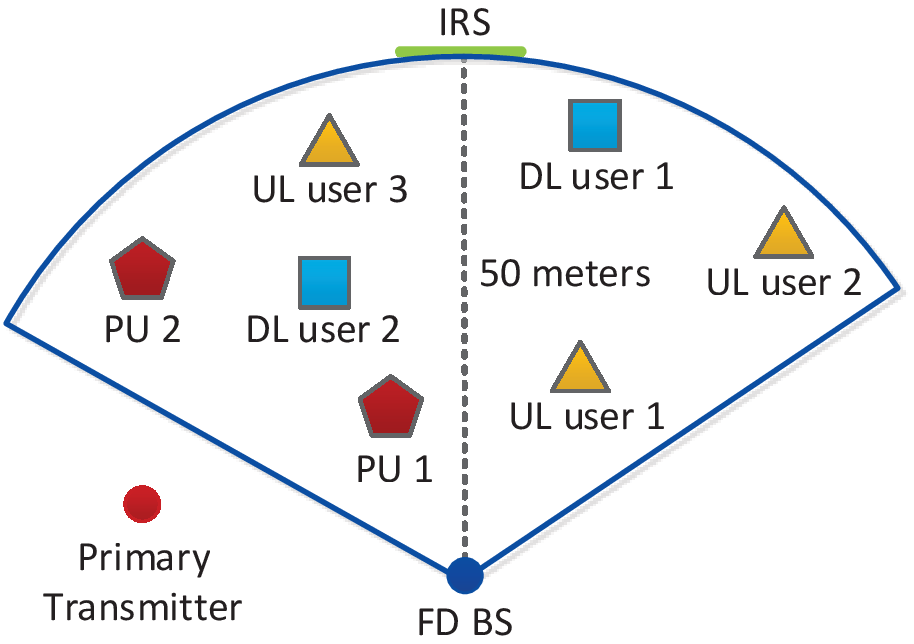}\vspace*{0mm}
\caption{Simulation setup for an IRS-assisted FD CR network which comprises $I=2$ PUs, $K=2$ DL users, and $J=3$ UL users.}
\label{simu_model}
\end{minipage}\hspace*{8mm}
\begin{minipage}[b]{0.47\linewidth} \hspace*{-1cm}
    \centering\includegraphics[width=3.4in]{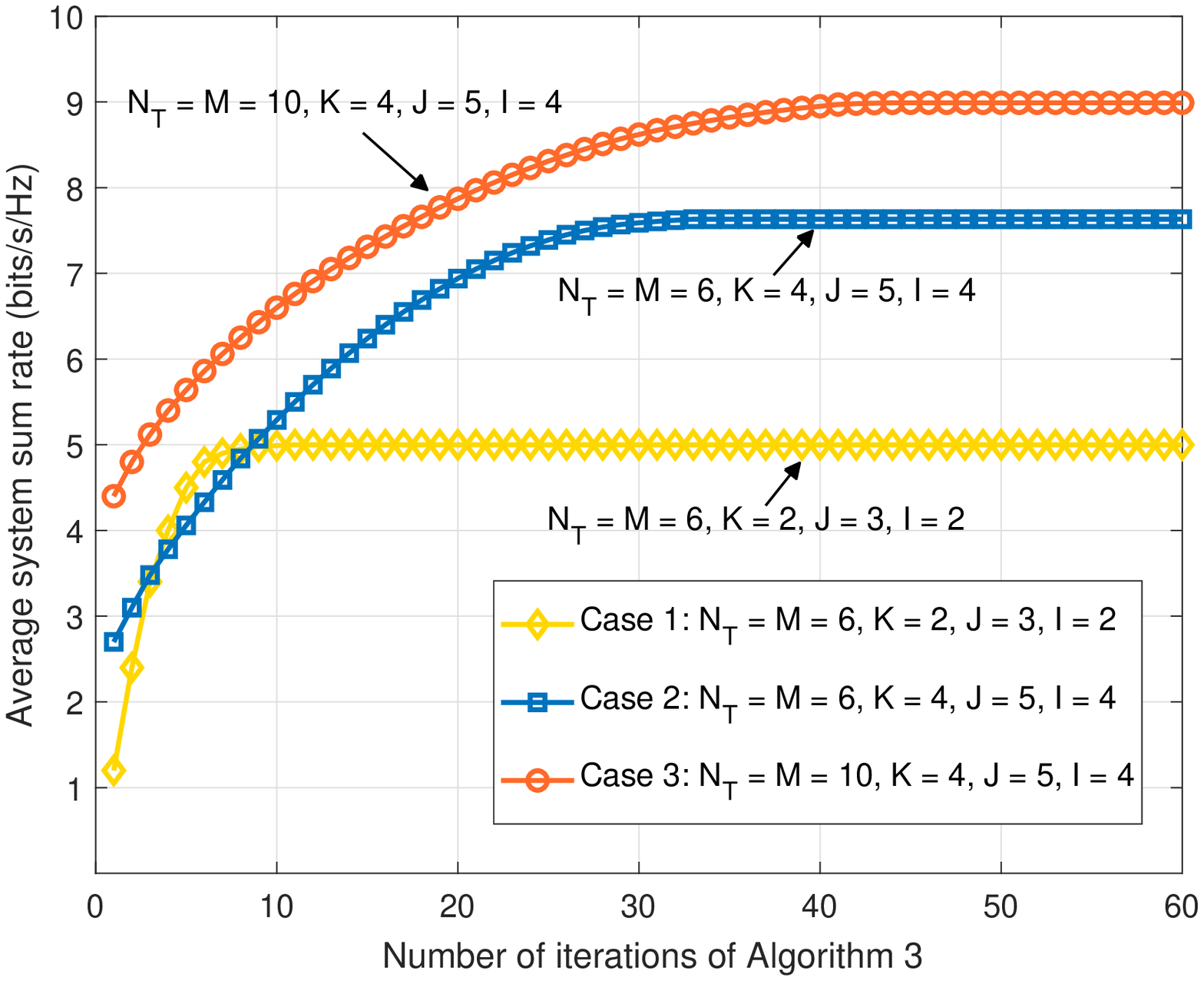}\vspace*{-7mm}
\caption{Convergence of the proposed BCD algorithm for different values of $N_{\mathrm{T}}$, $M$, $I$, $J$, and $K$ with $p_{\mathrm{tol}_i}=-90$ dBm, $\upsilon^2=10\%$, and $P^{\mathrm{DL}}_{\mathrm{max}}=30$ dBm.}
\label{convergence}
\end{minipage}\vspace*{-6mm}
\end{figure}

\begin{table}[t]\vspace*{0mm}\caption{System parameters adopted in simulations.}\vspace*{-4mm}\label{tab:parameters}\footnotesize
\newcommand{\tabincell}[2]{\begin{tabular}{@{}#1@{}}#2\end{tabular}}\vspace*{-0mm}
\centering
\begin{tabular}{|l|l|l||l|l|l|}\hline
    \hspace*{-1mm}$f_c$ & Carrier center frequency & $2.5$ GHz & $p_{j,\mathrm{max}}$ & Max. transmit power of UL user $j$ & $10$ dBm \\
\hline
    \hspace*{-1mm}$\alpha_{\mathrm{BU}}$ & Path loss exponent for direct paths & $3.9$ & $P^{\mathrm{DL}}_{\mathrm{max}}$ & Max. transmit power of FD BS & $30$ dBm  \\
\hline
    \hspace*{-1mm}$\alpha_{\mathrm{BR}}$ & Path loss exponent & $2.1$ & $\alpha_{\mathrm{RU}}$ & Path loss exponent & $2.3$  \\
\hline
    \hspace*{-1mm}$\eta$ & SI cancellation coefficient & $-85$ dB \cite{bharadia2013full} & $\sigma_{n_k}^2$ & Secondary DL user noise power  & $-100$ dBm \\
\hline
    \hspace*{-1mm}$\sigma_\mathrm{U}^2$ & Secondary FD BS noise power & $-110$ dBm & $G_i$ & BS antenna gain & $5$ dBi \\
\hline
    \hspace*{-1mm}$p_{\mathrm{tol}_i}$ & Interference tolerance & $90$ dBm & $G_{\mathrm{Rician}}$ & Rician factor for IRS channels & $5$ dB\\
\hline
    \hspace*{-1mm}$\varepsilon_{\mathrm{SCA}}$ & SCA error tolerance & $0.01$ & $\varepsilon_{\mathrm{BCD}}$ & BCD error tolerance & $0.01$ \\
\hline
    \hspace*{-1mm}$\chi$ & Penalty factor & $10^{3}$ & $\omega_j^{\mathrm{UL}}$, $\omega_k^{\mathrm{DL}}$ & Weights of UL and DL users & $1$ \\
\hline
\end{tabular}
\vspace*{0mm}
\end{table}
In this section, we study the system performance of the proposed resource allocation scheme via simulations. The schematic system model for the simulated FD CR network is shown in Figure \ref{simu_model}.
\subsection{Simulation Setup}
We focus on the resource allocation of one sector of the secondary network. The distance between the IRS and the secondary FD BS is 50 meter\footnote{ In practice, the location of the IRS can be either optimized or chosen for convenience.}. Unless specified otherwise, the primary network contains $I=2$ PUs while the secondary network comprises $K=2$ secondary DL users and $J=3$ secondary UL users. Both the PUs and the SUs are uniformly and randomly distributed in the considered sector. For the ease of presentation, in the sequel, the maximum normalized estimation errors of the PU CSI are defined as
$\upsilon^2_{\mathrm{D},i}=\frac{\varepsilon^2_{\mathrm{D},i}}{\left \| \mathbf{l}_{\mathrm{D},i} \right \|^2}$,
$\upsilon^2_{\mathrm{R},i}=\frac{\varepsilon^2_{\mathrm{R},i}}{\left \| \mathbf{l}_{\mathrm{R},i} \right \|^2}$, and $\upsilon^2_{i,j}=\frac{\varepsilon^2_{i,j}}{\left | e_{i,j}\right |^2}$, where $\upsilon_{\mathrm{D},i}^2=\upsilon^2_{\mathrm{R},i}=\upsilon^2_{i,j}=\upsilon^2$, $\forall i\in\mathcal{I}$, $\forall j\in\mathcal{J}$. Moreover, the path loss model for the reflected path is given by $PL_{\mathrm{R}}=c_{\mathrm{R}} (d_{\mathrm{BR}})^{-\alpha_{\mathrm{BR}}} (d_{\mathrm{RU}})^{-\alpha_{\mathrm{RU}}}$, where $c_{\mathrm{R}}=-80$ dB is a constant related to the carrier center frequency at the reference distance of 1 meter. Variables $d_{\mathrm{BR}}=50$ meter and $d_{\mathrm{RU}}$ are the distance between the FD BS and the IRS and the distance between the IRS and the users, respectively, and $\alpha_{\mathrm{BR}}=2.1$ and $\alpha_{\mathrm{RU}}=2.3$ are the corresponding path loss exponents \cite{wu2019intelligent}. On the other hand, the path loss model for the direct path is given by $PL_{\mathrm{D}}=c_{\mathrm{D}} (d_{\mathrm{BU}})^{-\alpha_{\mathrm{BU}}}$, where $c_{\mathrm{D}}=-40$ dB and the path loss exponent is $\alpha_{\mathrm{BU}}=3.9$\footnote{In practice, IRSs are usually deployed at favourable locations. As a result, we assume that the reflected signals suffer from a less severe path loss compared to the signals directly received from the BS.}. Besides, we model the multipath fading coefficients of the channels of the direct paths as independent and identically distributed Rayleigh random variables while the multipath fading coefficients of the channels of the reflected paths follow a Rician distribution. The adopted parameter values are listed in Table \ref{tab:parameters}.
\subsection{Baseline Schemes}
For comparison, we consider three baseline schemes. For baseline scheme 1, zero-forcing beamforming (ZF-BF) is employed at the FD BS for both DL and UL transmissions and the phases of the IRS are generated in a random manner. In particular, the directions of both the DL beamformer $\mathbf{w}_k$ for desired user $k$ and UL beamformer $\mathbf{v}_j$ for desired user $j$ are fixed and lie in the null spaces of all the other DL user channels and all the other UL user channels, respectively. Then, by optimizing the DL and UL transmit powers, i.e., $p_k^{\mathrm{DL}}\in\mathbb{R}$ and $p_j$, we solve the problem in \eqref{altprob1} subject to power constraints C1 and C2 and interference leakage constraint C4 by applying \textbf{Algorithm 3}\footnote{The optimization problem resulting for the baseline scheme 1 is still non-convex due to the coupling between the DL and UL transmit powers.}. For baseline scheme 2, we assume that the considered FD CR network does not employ an IRS\footnote{For baseline scheme 2, we solve \eqref{prob3} and \eqref{prob4} by applying \textbf{Algorithm 3} with  $\mathbf{\Psi}=\mathbf{0}$.}. Then, we optimize DL beamforming vectors $\mathbf{w}_k$, UL beamforming vectors $\mathbf{v}_j$, and the transmit powers of the UL users for maximization of the system sum rate subject to constraints C1, C2, and C4 in \eqref{prob1}. For baseline scheme 3, we assume that the secondary BS operates in the HD mode where the UL reception and the DL transmission are realized in two orthogonal time slots of equal duration. As a result, both CCI and SI do not exist. In particular, for the first time slot, the DL sum rate is maximized by jointly optimizing $\mathbf{w}_k$ and $\mathbf{\Psi}$ subject to constraints C1, C3, and C4 in \eqref{altprob1}\footnote{For maximization of the DL sum rate, we obtain $\mathbf{w}_k$ and $\mathbf{\Psi}$ by applying \textbf{Algorithm 3} with $p_j=0$ and $\mathbf{v}_j=\mathbf{0}$, $\forall j\in J$. Similarly, for maximization of the UL sum rate, we obtain $p_j$, $\mathbf{v}_j$, and $\mathbf{\Psi}$ by applying \textbf{Algorithm 3} with $\mathbf{w}_k=\mathbf{0}$, $\forall k\in K$.}. Then, for the second time slot, we maximize the UL sum rate by optimizing UL beamforming vector $\mathbf{v}_j$, phase shift matrix $\mathbf{\Psi}$, and the transmit powers of UL users, i.e., $p_j$ taking into account constraints C2, C3, and C4 in \eqref{prob1}. For a fair comparison, the resulting total sum rate obtained for baseline scheme 3 is multiplied by a factor of one half due to the orthogonal time slots needed for separating the UL and DL transmissions.
\subsection{Convergence of Algorithm 3}
In Figure \ref{convergence}, we investigate the convergence of the proposed BCD algorithm for different numbers of PUs $I$, secondary UL users $J$, secondary DL users $K$, antenna elements $N_{\mathrm{T}}$, and IRS reflecting elements $M$. In particular, we consider three cases: Case 1 with $N_{\mathrm{T}}=M=6$, $I=K=2$, and $J=3$; Case 2 with $N_{\mathrm{T}}=M=6$, $I=K=4$, and $J=5$; Case 3 with $N_{\mathrm{T}}=M=10$, $I=K=4$, and $J=5$. We can observe that for all three cases, the proposed algorithm monotonically converges to a stationary point. Specifically, for Case 1, the proposed algorithm converges within 10 iterations of \textbf{Algorithm 3}. For Case 2, the proposed algorithm needs considerably more iterations (roughly 30 iterations of \textbf{Algorithm 3}) to converge since the larger number of users leads to more optimization variables and constraints in \eqref{prob1}. Compared to Case 2, for Case 3, the proposed algorithm needs around 10 extra iterations for convergence since the larger values of $N_{\mathrm{T}}$ and $M$ enlarge the size of the solution space of the considered problem significantly. We also note that the number of iterations required for the proposed algorithm to converge is more sensitive to the number of users than to the number of antennas and reflecting elements.
\subsection{Average System Sum Rate versus Maximum DL Transmit Power}
\begin{figure}[t]
\centering
\begin{minipage}[b]{0.47\linewidth} \hspace*{-1cm}
\centering\includegraphics[width=3.4in]{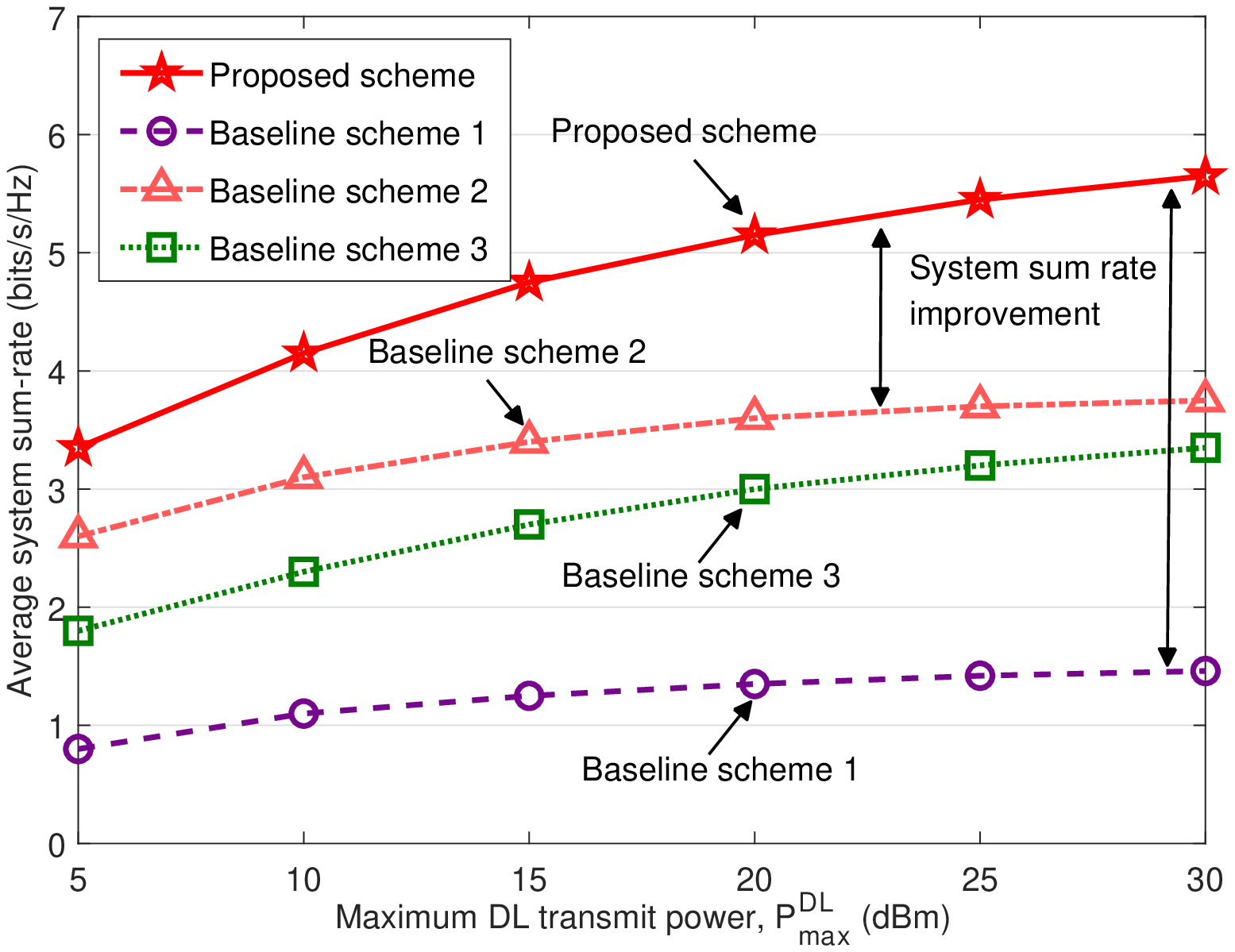}\vspace*{-7mm}
\caption{Average system sum rate (bits/s/Hz) versus maximum downlink transmit power (dBm) for different resource allocation schemes for $N_\mathrm{T}=8$, $M=8$, $I=2$, $J=3$, $K=2$, $p_{\mathrm{tol}_i}=-90$ dBm, and $\upsilon^2=10\%$.}
\label{srpower}
\end{minipage}\hspace*{8mm}
\begin{minipage}[b]{0.47\linewidth} \hspace*{-1cm}
    \centering\includegraphics[width=3.4in]{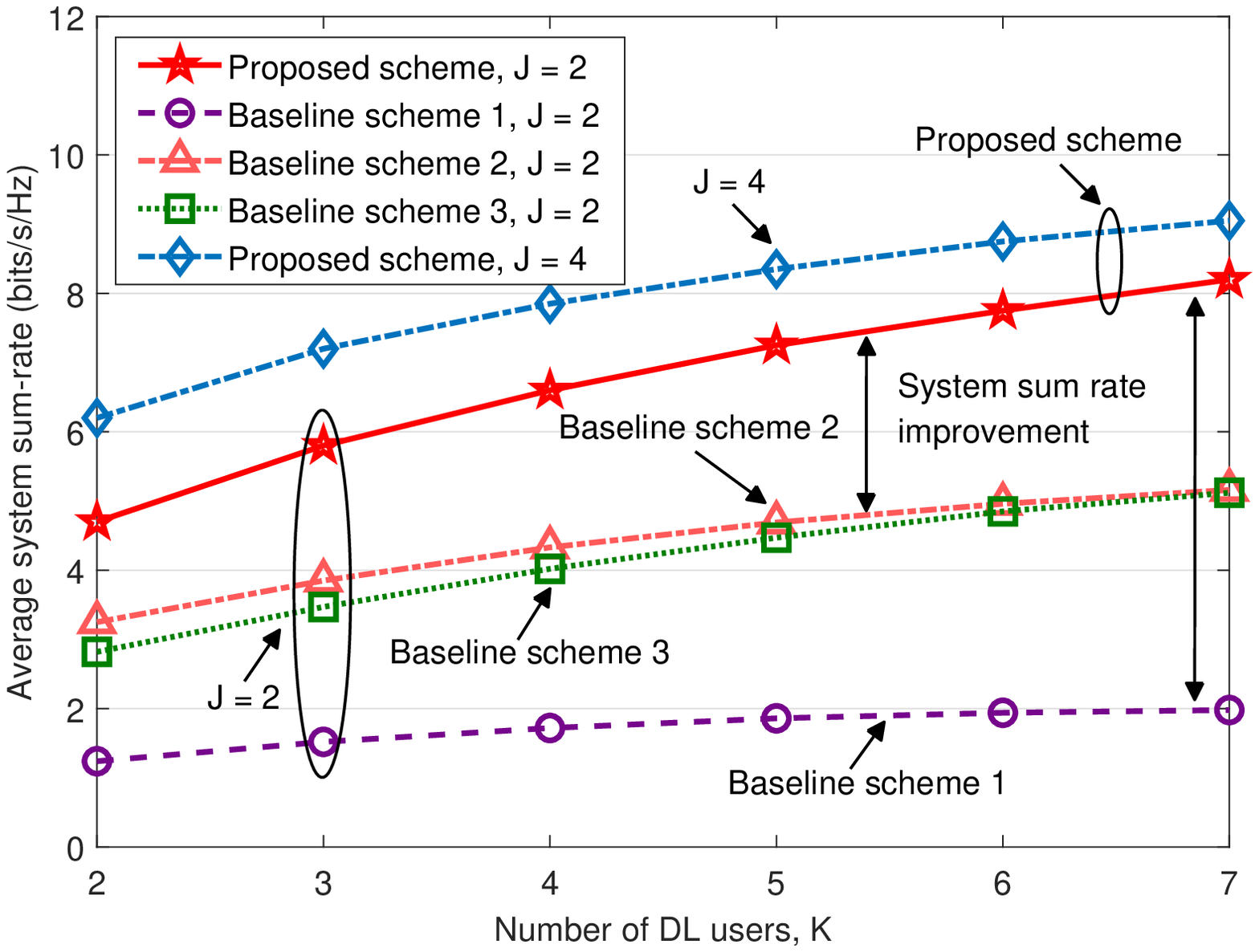}\vspace*{-7mm}
\caption{Average system sum rate (bits/s/Hz) versus maximum downlink transmit power (dBm) for different resource allocation schemes for $N_\mathrm{T}=8$, $M=8$, $I=2$, $\upsilon^2=10\%$, $p_{\mathrm{tol}_i}=-90$ dBm, and $P^{\mathrm{DL}}_{\mathrm{max}}=30$ dBm.}
\label{sruser}
\end{minipage}\vspace*{-6mm}
\end{figure}

In Figure \ref{srpower}, we study the average system sum rate versus the maximum DL transmit power, $P^{\mathrm{DL}}_{\mathrm{max}}$, for different resource allocation schemes. As expected, the system sum rate increases monotonically with $P^{\mathrm{DL}}_{\mathrm{max}}$. Moreover, we observe that the proposed scheme outperforms all baseline schemes. In fact, compared to the baseline schemes, the significant performance improvement achieved by the proposed resource allocation scheme is enabled by the joint optimization of $\mathbf{\Phi}$, $\mathbf{w}_k$, $p_j$, and $\mathbf{v}_j$. On the one hand, the proposed scheme can create a more favorable radio propagation environment by optimizing the phase shift matrix of the IRS. On the other hand, it can fully exploit the DoFs introduced by the multiplexing of multiple UL and DL users on the same spectral resource via FD, which improves the spectral efficiency of the CR network. On the contrary, the three baseline schemes yield dramatically lower system sum rates. Specifically, for baseline scheme 1, the FD BS is unable to fully exploit the DoFs available for resource allocation because of the fixed beamforming vector. Although the multiuser interference (MUI) is mitigated by ZF-BF, both the CCI and the remaining SI become more serious as $P^{\mathrm{DL}}_{\mathrm{max}}$ increases which limits the system sum rate. For baseline scheme 2, since there is no IRS available, there are no DoFs for customizing a favorable radio propagation environment for enhancing the desired signal and suppressing the interference at the PUs. For baseline scheme 3, although orthogonal DL and UL transmissions completely avoid CCI and SI, the resulting strictly suboptimal use of the DL and UL time resources leads to a significant loss of spectral efficiency.
\subsection{Average System Sum Rate versus Number of Secondary Users}
Figure \ref{sruser} depicts the average system sum rate versus the number of secondary DL users for different resource allocation schemes. As can be seen from Figure \ref{sruser}, as $K$ grows, the system sum rates achieved with the proposed scheme and the three baseline schemes increase since all schemes are able to exploit multiuser diversity. Similarly, we observe that the performance of the proposed scheme improves when the number of UL users, $J$, increases. However, compared to the proposed scheme, the system sum rates for the baseline schemes are significantly lower. In particular, due to the partially fixed beamforming pattern of baseline scheme 1, the increasing CCI and SI associated with larger $K$ cannot be mitigated which results in a substantially lower system sum rate. For baseline scheme 2, since the IRS is not utilized, the system is unable to mitigate the growing MUI in UL and DL introduced by the increasing number of DL users $K$. For baseline scheme 3, the achieved system sum rate is still lower compared to the proposed scheme due to the inefficient utilization of radio spectrum caused by the HD BS.
\begin{figure}[t]
\centering
\begin{minipage}[b]{0.47\linewidth} \hspace*{-1cm}
\centering\includegraphics[width=3.4in]{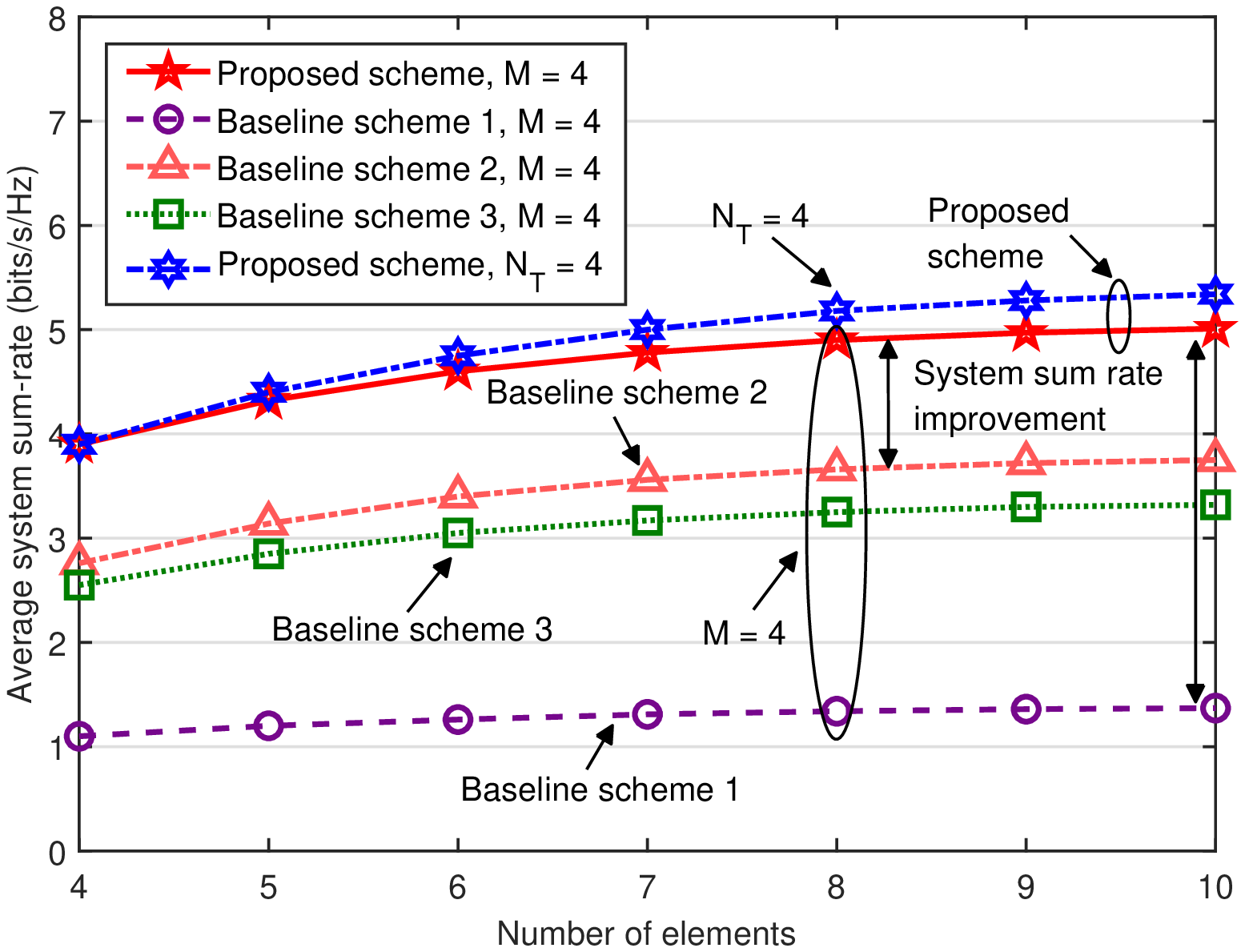}\vspace*{-7mm}
\caption{Average system sum rate (bits/s/Hz) versus number of elements for different resource allocation schemes for $K=2$, $I=2$, $J=3$, $\upsilon^2=10\%$, $p_{\mathrm{tol}_i}=-90$ dBm, and $P^{\mathrm{DL}}_{\mathrm{max}}=30$ dBm.}
\label{srelement}
\end{minipage}\hspace*{8mm}
\begin{minipage}[b]{0.47\linewidth} \hspace*{-1cm}
    \centering\includegraphics[width=3.4in]{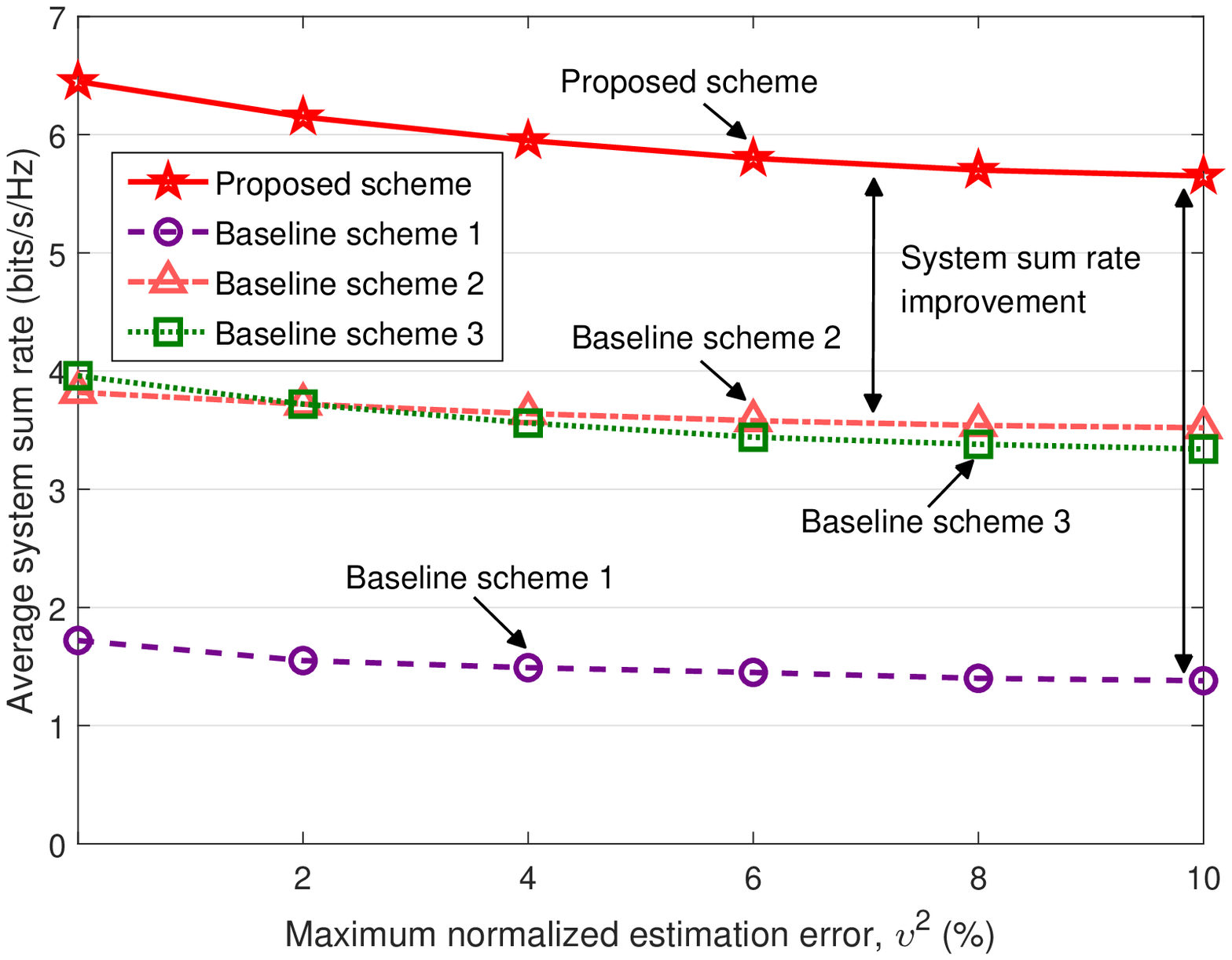}\vspace*{-7mm}
\caption{Average system sum rate (bits/s/Hz) versus maximum normalized channel estimation error, $\upsilon^2$, for different schemes for $N_\mathrm{T}=M=8$, $K=I=2$, $J=3$, $p_{\mathrm{tol}_i}=-90$ dBm, and $P^{\mathrm{DL}}_{\mathrm{max}}=30$ dBm.}
\label{sruncertainty}
\end{minipage}\vspace*{-6mm}
\end{figure}
\subsection{Average System Sum Rate versus Number of Antenna/IRS Elements}
In Figure \ref{srelement}, we investigate the average system sum rate versus the number of antenna/IRS elements. Specifically, to reveal the performance gain achieved by deploying an IRS, for the proposed scheme two cases are considered: Case 1 with a fixed number of antennas at the secondary BS ($N_{\mathrm{T}}=4$) and increasing $M$ and Case 2 with a fixed number of phase shifters ($M=4$) and increasing $N_{\mathrm{T}}$. We observe that increasing the number of elements in Case 1 results in a larger performance gain compared to Case 2. The reason behind this is twofold. On the one hand, as the number of reflectors at the IRS increases, there are more DoFs for customizing favorable BS-IRS-user channels which improves both the UL and the DL beamforming gain. On the other hand, the additional IRS elements can reflect more power of the signal transmitted by the secondary FD BS which results in a power gain. Moreover, as can be seen from Figure \ref{srelement}, the average system sum rates for the proposed scheme and the three baseline schemes improve as the number of antennas, $N_{\mathrm{T}}$, at the FD BS increases. This can be explained by the fact that the extra DoFs provided by the additional antennas facilitates a higher beamforming resolution for both DL transmission and UL reception which lead to higher received SINRs. Yet, as $N_{\mathrm{T}}$ increases, the channel hardening effect leads to a diminishing growth rate of the system sum rate. Figure \ref{srelement} also shows that the average system sum rate of the proposed scheme increases faster with $N_{\mathrm{T}}$ than the average system sum rates
of the baseline schemes thanks to the proposed optimization framework which exploits the system resources efficiently. 
\subsection{Average System Sum Rate versus Maximum Normalized Channel Estimation Error}
In Figure \ref{sruncertainty}, we study the average system sum rate versus the maximum normalized channel estimation error. As expected, the average system sum rate decreases with increasing $\upsilon^2$. This is due to the fact that, as $\upsilon^2$ increases, the secondary
BS becomes less flexible and more conservative in resource allocation. In particular, the BS has to allocate more DoFs to satisfy the interference leakage constraint C4. As a result, fewer DoFs are available for suppressing the SI and facilitating accurate DL beamforming at the FD BS which degrades the system performance. Besides, over the entire range of $\upsilon^2$, the proposed scheme significantly outperforms the three baseline schemes. This unveils that by jointly optimizing all available DoFs, the proposed scheme can mitigate the interference leakage more efficiently than the three baseline schemes, even in the presence of CSI uncertainty. Besides, compared to the proposed scheme and baseline scheme 3, we observe that baseline scheme 1 and baseline 2 are less sensitive to channel estimation errors in the considered range. For baseline scheme 1, the random phase shift pattern of the IRS already results in a significant performance loss and increasing $\upsilon^2$ from 0 to 10$\%$ only leads to a small additional loss. For baseline scheme 2, since the IRS is not deployed, only the imperfect knowledge of the CSI of the direct paths affects the performance, which leads to a smaller degradation. 
\begin{figure}[t]
    \centering\includegraphics[width=3.4in]{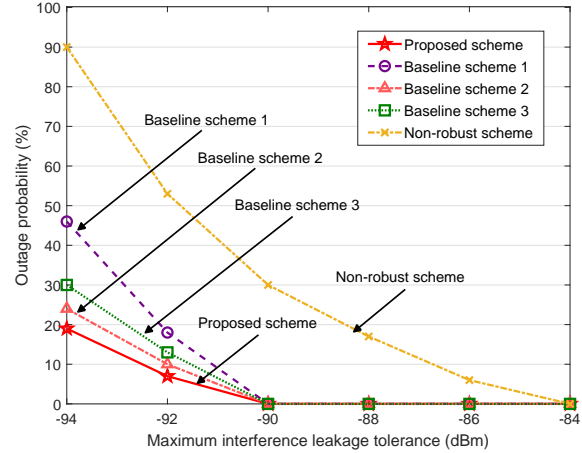}\vspace*{-7mm}
\caption{Outage probability (\%) versus maximum interference leakage tolerance (dBm) for different resource allocation schemes for $N_\mathrm{T}=8$, $M=8$, $K=2$, $I=2$, $J=3$, $\upsilon^2=10\%$ , and $P^{\mathrm{DL}}_{\mathrm{max}}=30$ dBm.}
\label{probinterference}\vspace*{-6mm}
\end{figure}
\subsection{Outage Probability versus Maximum Interference Leakage Tolerance}
Figure \ref{probinterference} shows the outage probability of the users in the primary network versus the maximum interference leakage tolerance for different resource allocation schemes. The outage probability is defined as the probability that the interference leakage from the secondary network to the $i$-th PU is higher than a predefined target interference leakage tolerance $p_{\mathrm{tar}_i}$. For comparison, we also study the outage probability of a non-robust scheme for comparison. Specifically, for the non-robust scheme, we solve a problem similar to \eqref{prob1} but treat the estimated CSI of the PUs as perfect CSI. Then, using the actual CSI of the PUs, we check if the interference leakage constraint C4 in \eqref{prob1} is satisfied. As can be observed from Figure \ref{probinterference}, both the proposed scheme and the baseline schemes yield a significant outage probability reduction compared to the non-robust scheme. Moreover, as we set the maximum interference leakage tolerance to $p_{\mathrm{tol}_i}=-90$ dBm, the outage probabilities of the proposed scheme and all baseline schemes decrease to zero for target interference leakage tolerances $p_{\mathrm{tar}_i}\leq -90$ dBm. In contrast, the non-robust scheme still suffers from outages. These results underline the robustness of the proposed scheme against imperfect CSI.
\section{Conclusion}
In this paper, we proposed to integrate an IRS into a multiuser FD CR system to simultaneously improve the system performance of the secondary network and effectively mitigate the interference caused to the PUs. In particular, the system sum rate of the secondary network was maximized by jointly optimizing the DL transmit beamforming vectors and the UL receive beamforming vectors at the FD BS, the UL transmit power of the UL users, and the phase shift matrix at the IRS. We considered the robust design of IRS-assisted FD CR systems taking into account the imperfect knowledge of the CSI of the PUs. Since the resulting interference leakage tolerance constraint is an obstacle to efficient resource allocation algorithm design, we proposed a safe approximation of the original optimization problem. To tackle the non-convexity of the resulting design problem, we developed a BCD algorithm to solve the approximated problem in an alternating manner. In particular, the design of the DL transmit beamformers and UL transmit power was tackled by SCA and SDR, and the optimal UL receive beamformers were derived in closed form. The unit modulus constrained optimization problem introduced by the IRS was first transformed to a rank-constrained problem and then solved by applying a penalty method and SCA. The proposed BCD algorithm is guaranteed to converge to a stationary point of the approximated optimization problem. Simulation results not only revealed the significant system sum rate improvement achieved by the proposed scheme compared to three baseline schemes but also verified its robustness against the imperfect knowledge of the CSI of the PUs. Moreover, our results illustrated that IRSs are an efficient means to mitigate the various forms of interference in FD CR systems.
\section*{Appendix}
\subsection{Proof of Theorem 1}
To start with, we recast the relaxed version of \eqref{prob3} in equivalent form as follows
\begin{eqnarray}
\label{prob11}
&&\hspace*{-22mm}\underset{\substack{\mathbf{W}_k,p_j,\beta_i,\gamma_i,\tau_i,\\\delta_i,\iota_i,\kappa_i,\widehat{\phi }_k,\widehat{\psi }_j}}{\mino} \,\, \,\, -\underset{ k\in\mathcal{K}}{\sum}\mathrm{log}_2(\widehat{\phi }_k+\sigma^2_{\mathrm{n}_k})-\underset{ j\in\mathcal{J}}{\sum}\mathrm{log}_2(\widehat{\psi }_j)-\underset{k\in\mathcal{K}}{\sum}\mathrm{Tr}\Big(\big(\nabla_{\mathbf{W}_k}\widehat{g}_1+\nabla_{\mathbf{W}_k}\widehat{g}_2\big)^H\mathbf{W}_k\Big)+\Xi\notag\\
\mbox{s.t.}\hspace*{-2mm}
&&\mbox{C1},\mbox{C2},\widehat{\mbox{C4}}\mbox{a},\widehat{\mbox{C4}}\mbox{b},\widehat{\mbox{C4}}\mbox{c},\widehat{\mbox{C4}}\mbox{d},\mbox{C5},\notag\\
&&\mbox{C10}:\widehat{\phi }_k\leq\underset{r\in\mathcal{K} }{\sum }\mathrm{Tr}(\widehat{\mathbf{g}}_k\widehat{\mathbf{g}}_k^H\mathbf{W}_r)+\underset{j\in\mathcal{J}}{\sum}p_j \left | \varphi_{j,k}\right |^2,~\forall k,\notag\\
&&\mbox{C11}:\widehat{\psi }_j\leq\underset{t\in\mathcal{J} }{\sum }p_t\mathrm{Tr}(\widehat{\mathbf{h}}_j\widehat{\mathbf{h}}_j^H\mathbf{v}_j\mathbf{v}_j^H)+\mathrm{Tr}\big (\eta\mathbf{v}_j\mathbf{v}_j^H\mathrm{Diag}\big (\underset{k\in\mathcal{K}}{\sum }\mathbf{S}\mathbf{W}_k\mathbf{S}^H\big )\big )+\sigma^2_{\mathrm{U}}\left \| \mathbf{v}_j \right \|^2,~\forall j,
\end{eqnarray}
where $\widehat{\phi }_k$ and $\widehat{\psi }_j$ are slack variables and $\Xi$ collects all terms which are not relevant for the proof. Note that the problem in \eqref{prob11} is jointly convex with respect to all optimization variables and the Slater's
condition is satisfied for \eqref{prob11} \cite{boyd2004convex}. Therefore, strong duality holds, i.e., the gap between \eqref{prob11} and its dual problem is zero \cite{boyd2004convex}. Specifically, the Lagrangian function of \eqref{prob11} in terms of beamforming matrix $\mathbf{W}_\mathit{k}$ is given as follows
\vspace*{-2mm}
\begin{eqnarray}
\label{Lagrangian}
\mathcal{L}&&\hspace*{-6mm}=-\underset{k\in\mathcal{K}}{\sum}\mathrm{Tr}\Big(\big(\nabla_{\mathbf{W}_k}\widehat{g}_1+\nabla_{\mathbf{W}_k}\widehat{g}_2\big)^H\mathbf{W}_k\Big)+\xi\underset{k\in\mathcal{K}}{\sum }\mathrm{Tr}(\mathbf{W}_k)-\underset{i\in\mathcal{I}}{\sum }\mathrm{Tr}\big(\mathbf{S}_{\widehat{\mathrm{C}4}\mathrm{b}_{i}}(\mathbf{W_\mathit{k}},\beta_i,\gamma_i,\kappa_i)\mathbf{T}_{\widehat{\mathrm{C}4}\mathrm{b}_{i}}\big)\notag\\&&\hspace*{-6mm}-\underset{i\in\mathcal{I}}{\sum }\mathrm{Tr}\big(\mathbf{S}_{\widehat{\mathrm{C}4}\mathrm{c}_{i}}(\mathbf{W_\mathit{k}},p_j,\bm{\Psi},\gamma_i,\tau_i,\delta_i)\mathbf{T}_{\widehat{\mathrm{C}4}\mathrm{c}_{i}}\big)+\underset{i\in\mathcal{I}}{\sum }\widehat{\beta}_i\underset{r\in\mathcal{K} }{\sum }\mathrm{Tr}(\widehat{\mathbf{l}}_i\widehat{\mathbf{l}}_i^H\mathbf{W}_r)-\underset{k\in\mathcal{K}}{\sum }\widehat{\zeta}_k\underset{r\in\mathcal{K} }{\sum }\mathrm{Tr}(\widehat{\mathbf{g}}_k\widehat{\mathbf{g}}_k^H\mathbf{W}_r)
\notag\\
&&\hspace*{-6mm}-\underset{j\in\mathcal{J}}{\sum }\widehat{\varsigma} _j\mathrm{Tr}\big (\eta\mathbf{v}_j\mathbf{v}_j^H\mathrm{Diag}\big (\underset{k\in\mathcal{K}}{\sum }\mathbf{S}\mathbf{W}_k\mathbf{S}^H\big )\big )-\underset{k\in\mathcal{K}}{\sum}\mathrm{Tr}(\mathbf{W}_\mathit{k}\mathbf{Y}_\mathit{k})+\Upsilon.
\end{eqnarray}
Here, we have introduced $\Upsilon$ to collect all terms that do not involve $\mathbf{W}_\mathit{k}$. The scalar Lagrange multipliers $\xi$, $\widehat{\beta}_i$, $\widehat{\zeta}_k$, and $\widehat{\varsigma}_j\geq 0$ are associated with constraints $\mbox{C1}$, $\widehat{\mbox{C4}}\mbox{d}$, $\mbox{C10}$, and $\mbox{C11}$, respectively. The positive semidefinite Lagrange multiplier matrices $\mathbf{T}_{\widehat{\mathrm{C}4}\mathrm{b}_{i}}\in \mathbb{C}^{(N_\mathrm{T}+1)\times (N_\mathrm{T}+1)}$, $\mathbf{T}_{\widehat{\mathrm{C}4}\mathrm{c}_{i}}\in \mathbb{C}^{(M+1)\times (M+1)}$, and $\mathbf{Y}_k\in \mathbb{C}^{N_\mathrm{T}\times N_\mathrm{T}}$ are associated with constraints $\widehat{\mbox{C4}}\mbox{b}$, $\widehat{\mbox{C4}}\mbox{c}$, and $\mbox{C5}$, respectively. The dual problem of \eqref{prob11} is given by
\vspace*{-1mm}
\begin{equation}
\underset{\substack{\mathbf{T}_{\widehat{\mathrm{C}4}\mathrm{b}_{i}},\mathbf{T}_{\widehat{\mathrm{C}4}\mathrm{c}_{i}},\mathbf{Y}_k\succeq \mathbf{0},\\\xi,\widehat{\beta}_i,\widehat{\zeta}_k,\widehat{\varsigma}_j\geq0~}}{\maxo}~~ \underset{\substack{\mathbf{W}_k,p_j,\beta_i,\gamma_i,\tau_i,\\\delta_i,\iota_i,\kappa_i,\widehat{\phi }_k,\widehat{\psi }_j}}{\mino}~~ \mathcal{L}(\mathbf{W}_k,p_j,\beta_i,\gamma_i,\tau_i,\delta_i,\iota_i,\kappa_i,\widehat{\phi}_k,\widehat{\psi }_j,\mathbf{T}_{\widehat{\mathrm{C}4}\mathrm{b}_{i}},\mathbf{T}_{\widehat{\mathrm{C}4}\mathrm{c}_{i}},\mathbf{Y}_k,\xi,\widehat{\beta}_i,\widehat{\zeta}_k,\widehat{\varsigma}_j).\label{DP}\\[-1mm]
\end{equation}
Next, by checking the Karush-Kuhn-Tucker (KKT) conditions with respect to $\mathbf{W}_k$, we investigate the structure of the optimal $\mathbf{W}^{\dagger}_k$ of \eqref{DP}. Specifically, for $\mathbf{W}^{\dagger}_k$, we have
\vspace*{-2mm}
\begin{equation}
\label{KKTCond}
\mbox{K1:~}\xi^{\dagger},\widehat{\beta}_i^{\dagger},\widehat{\zeta}_k^{\dagger},\widehat{\varsigma}_j^{\dagger}\geq 0,~ \mathbf{T}^{\dagger}_{\widehat{\mathrm{C}4}\mathrm{b}_{i}},\mathbf{T}^{\dagger}_{\widehat{\mathrm{C}4}\mathrm{c}_{i}},\mathbf{Y}_k^{\dagger}\succeq \mathbf{0},~~~
\mbox{K2:~}\mathbf{Y}_k^{\dagger}\mathbf{W}_k^{\dagger}=\mathbf{0},~~~
\mbox{K3:~}\nabla_{\mathbf{W}_k^{\dagger}}\mathcal{L}=\mathbf{0},
\end{equation}
where $\xi^{\dagger}$, $\widehat{\beta}_i^{\dagger}$, $\zeta_c^{\dagger}$, $\mathbf{T}^{\dagger}_{\widehat{\mathrm{C}4}\mathrm{b}_{i}}$, $\mathbf{T}^{\dagger}_{\widehat{\mathrm{C}4}\mathrm{c}_{i}}$, and $\mathbf{Y}_k^{\dagger}$ are the optimal Lagrange multipliers for \eqref{DP}. Note that there exists at least one $\xi^{\dagger}>0$ since constraint C1 is active for optimal $\mathbf{W}^{\dagger}_k$. To facilitate the proof, $\mbox{K3}$ in \eqref{KKTCond} is explicitly expressed as follows
\vspace*{-2mm}
\begin{equation}
    \mathbf{Y}_k^{\dagger}=\xi^{\dagger}\mathbf{I}_{N_{\mathrm{T}}}-\mathbf{\Delta}^{\dagger}_k,\label{alK3}\\[-2mm]
\end{equation} 
where $\mathbf{\Delta}_k^{\dagger}$ is given by
\begin{eqnarray}
\mathbf{\Delta}_k^{\dagger}&&\hspace*{-5mm}=\nabla_{\mathbf{W}_k}\widehat{g}_1+\nabla_{\mathbf{W}_k}\widehat{g}_2-\underset{i\in\mathcal{I} }{\sum }\mathbf{E}^H_i\mathbf{T}^{\dagger}_{\widehat{\mathrm{C}4}\mathrm{b}_{i}}\mathbf{E}_i-\underset{i\in\mathcal{I} }{\sum }\mathbf{C}^H\mathbf{\Psi}\mathbf{F}\mathbf{T}^{\dagger}_{\widehat{\mathrm{C}4}\mathrm{c}_{i}}\mathbf{F}^H\mathbf{\Psi}^H\mathbf{C}\notag\\&&\hspace*{-5mm}-\underset{i\in\mathcal{I}}{\sum }\widehat{\beta}_i\widehat{\mathbf{l}}_i\widehat{\mathbf{l}}_i^H+\underset{k\in\mathcal{K}}{\sum }\widehat{\zeta}_k\widehat{\mathbf{g}}_k\widehat{\mathbf{g}}_k^H+\underset{j\in\mathcal{J}}{\sum }\widehat{\varsigma} _j\eta\mathbf{v}_j\mathbf{v}_j^H\mathrm{Diag}(\mathbf{S}\mathbf{S}^H).
\end{eqnarray}
\par
Next, by unveiling the structure of matrix $\mathbf{Y}_k^{\dagger}$, we show that the optimal $\mathbf{W}^{\dagger}$ always satisfies $\mathrm{Rank}(\mathbf{W}_k^{\dagger})\leq 1$. Denote the maximum eigenvalue of matrix $\mathbf{\Delta}_k^{\dagger}$ as $\nu ^{\mathrm{max}}_{\mathbf{\Delta}_k^{\dagger}}\in\mathbb{R}$. We note
that due to the randomness of
the channels, the probability of having multiple eigenvalues with the same value
$\nu ^{\mathrm{max}}_{\mathbf{\Delta}_k^{\dagger}}$ is zero. Reviewing \eqref{alK3}, if $\nu ^{\mathrm{max}}_{\mathbf{\Delta}_k^{\dagger}}>\xi^{\dagger}$, then $\mathbf{Y}_k^{\dagger}\succeq\mathbf{0}$ does not hold which contradicts K1. On the other hand, if $\nu ^{\mathrm{max}}_{\mathbf{\Delta}_k^{\dagger}}\leq\xi^{\dagger}$, then $\mathbf{Y}_k^{\dagger}$ is a positive semidefinite matrix with $\mathrm{Rank}(\mathbf{Y}_k^*)\geq N_\mathrm{T}-1$. Considering K2, this leads to $\mathrm{Rank}(\mathbf{W}_k^{\dagger})\leq 1$. Next, we construct a bounded optimal solution based on the above discussion. Specifically, we construct a unit-norm vector $\mathbf{e}^{\mathrm{max}}_{\mathbf{\Delta}_k^{\dagger}}\in \mathbb{C}^{N_{\mathrm{T}}\times 1}$ which lies in the null space of $\mathbf{Y}_k^{\dagger}$, i.e., $\mathbf{Y}_k^{\dagger}\mathbf{e}^{\mathrm{max}}_{\mathbf{\Delta}_k^{\dagger}}=\mathbf{0}$. Let $\mathbf{e}^{\mathrm{max}}_{\mathbf{\Delta}_k^{\dagger}}$ be the unit-norm eigenvector associated with the principal eigenvalue $\nu ^{\mathrm{max}}_{\mathbf{\Delta}_k^{\dagger}}$ of matrix $\mathbf{\Delta}_k^{\dagger}$. Thus, the optimal $\mathbf{W}_k^*$ can be expressed as $\mathbf{W}_k^{\dagger}= \varpi \mathbf{e}^{\mathrm{max}}_{\mathbf{\Delta}_k^{\dagger}}(\mathbf{e}^{\mathrm{max}}_{\mathbf{\Delta}_k^{\dagger}})^H$. Here, parameter $\varpi$ can be tuned such that the DL transmit power constraint $\mbox{C1}$ is satisfied. \qed
\subsection{Proof of Theorem 2}
To start with, we define the objective function and the optimal solution of problem \eqref{prob7} as $\widetilde{F}(\mathbf{\Theta})$ and $\mathbf{\Theta}^{\dagger}$, respectively. Then, for any feasible $\mathbf{\Theta}$, we have the following inequality:
\begin{equation}
\label{wideF}
    \widetilde{F}(\mathbf{\Theta}^{\dagger})\leq\widetilde{F}(\mathbf{\Theta}).
\end{equation}
We further define the objective function of problem \eqref{prob8} as $\widetilde{G}(\mathbf{\Theta};\chi)$. Assuming $\mathbf{\Theta}_q$ minimizes $\widetilde{G}(\cdot;\chi_q)$ with penalty factor $\chi_q$ for each $q$, we have the following inequality:
\begin{eqnarray}
\label{Ginequality}
&&\hspace*{-6mm}\widetilde{F}(\mathbf{\Theta}_q)+\chi_q\big(\left \|\mathbf{\Theta}_q\right \|_*-\left \| \mathbf{\Theta}_q\right \|_2)=\widetilde{G}(\mathbf{\Theta}_q;\chi_q)\notag\\\leq
&&\hspace*{-6mm}\widetilde{F}(\mathbf{\Theta}^{\dagger})+\chi_q(\left \|\bm{\Theta}^{\dagger}\right \|_*-\left \| \mathbf{\Theta }^{\dagger}\right \|_2)=\widetilde{G}(\mathbf{\Theta}^{\dagger};\chi_q)\overset{(a)}{=}\widetilde{F}(\mathbf{\Theta}^{\dagger}),
\end{eqnarray}
where equality $(a)$ holds due to the fact that any optimal solution of \eqref{prob7}, i.e., $\bm{\Theta}^{\dagger}$, fulfills $\left \|\bm{\Theta}^{\dagger}\right \|_*-\left \| \mathbf{\Theta }^{\dagger}\right \|_2\leq 0$. Then, we rearrange the inequality in \eqref{Ginequality} and obtain the following inequality:
\begin{equation}
\label{Finequality}
    \left \|\bm{\Theta}_q\right \|_*-\left \| \mathbf{\Theta }_q\right \|_2\leq\frac{1}{\chi_q}\big(\widetilde{F}(\mathbf{\Theta}^{\dagger})-\widetilde{F}(\mathbf{\Theta}_q)\big).
\end{equation}
Recall that if $\overline{\bm{\Theta}}$ is a limit point of the sequence $\left \{ \bm{\Theta}_q \right \}$, we can find an infinite subsequence  $\mathcal{Q}$ such that 
\begin{equation}
    \lim_{q\in\mathcal{Q}}\mathbf{\Theta}_q=\overline{\mathbf{\Theta}}.
\end{equation}
Then, as $q\in\mathcal{Q}$, $q\rightarrow\infty$, we take the limit on both sides of \eqref{Finequality} and obtain the following relation chain:
\begin{equation}
    \left \|\overline{\mathbf{\Theta}}\right \|_*-\left \| \overline{\mathbf{\Theta}}\right \|_2\overset{(b)}{=}\lim_{q\in\mathcal{Q}}\left \|\mathbf{\Theta}_q\right \|_*-\left \| \mathbf{\Theta }_q\right \|_2\leq\lim_{q\in\mathcal{Q}}\frac{1}{\chi_q}\big(\widetilde{F}(\mathbf{\Theta}^{\dagger})-\widetilde{F}(\mathbf{\Theta}_q)\big)\overset{(c)}{=}0,
\end{equation}
where equality $(b)$ holds because of the continuity property of norm functions and equality $(c)$ holds due to $\chi_q\rightarrow\infty$. Thus, we have that $\left \|\overline{\mathbf{\Theta}}\right \|_*-\left \| \overline{\mathbf{\Theta}}\right \|_2=0$. As a result, $\overline{\mathbf{\Theta}}$ is a feasible solution of problem \eqref{prob7}.
\par
On the other hand, for any $\chi_q\geq 0$, we take the limit of \eqref{Ginequality} as $i\in\mathcal{Q}$, $q\rightarrow\infty$, which leads to the inequality:
\begin{equation}
\label{Fbarinequality}
    \widetilde{F}(\overline{\mathbf{\Theta}})\overset{(d)}{\leq}\widetilde{F}(\overline{\mathbf{\Theta}})+\lim_{q\in\mathcal{Q}}\chi_q(\left \|\bm{\Theta}_q\right \|_*-\left \| \mathbf{\Theta }_q\right \|_2)\leq\widetilde{F}(\mathbf{\Theta}^{\dagger}),
\end{equation}
where inequality $(d)$ is due to the nonnegativity of the term $\left \|\bm{\Theta}_q\right \|_*-\left \| \mathbf{\Theta }_q\right \|_2$. As $\overline{\mathbf{\Theta}}$ is a feasible point whose objective value is no larger than that of the optimal solution $\mathbf{\Theta }^{\dagger}$, we conclude that $\overline{\mathbf{\Theta}}$ is also an optimal solution of problem \eqref{prob7}, as claimed. This completes the proof.
\vspace*{-1mm}
\bibliographystyle{IEEEtran}
\bibliography{IRS_journal_reference_list}

\begin{thebibliography}{10}
\providecommand{\url}[1]{#1}
\csname url@samestyle\endcsname
\providecommand{\newblock}{\relax}
\providecommand{\bibinfo}[2]{#2}
\providecommand{\BIBentrySTDinterwordspacing}{\spaceskip=0pt\relax}
\providecommand{\BIBentryALTinterwordstretchfactor}{4}
\providecommand{\BIBentryALTinterwordspacing}{\spaceskip=\fontdimen2\font plus
\BIBentryALTinterwordstretchfactor\fontdimen3\font minus
  \fontdimen4\font\relax}
\providecommand{\BIBforeignlanguage}[2]{{%
\expandafter\ifx\csname l@#1\endcsname\relax
\typeout{** WARNING: IEEEtran.bst: No hyphenation pattern has been}%
\typeout{** loaded for the language `#1'. Using the pattern for}%
\typeout{** the default language instead.}%
\else
\language=\csname l@#1\endcsname
\fi
#2}}
\providecommand{\BIBdecl}{\relax}
\BIBdecl

\bibitem{xu2020resource}
D.~Xu, X.~Yu, and R.~Schober, ``Resource allocation for intelligent reflecting
  surface-assisted cognitive radio networks,'' \emph{submitted to SPAWC 2020,
  arXiv preprint arXiv:2001.11729}, 2020.

\bibitem{wong2017key}
V.~W. Wong, R.~Schober, D.~W.~K. Ng, and L.-C. Wang, \emph{Key Technologies for
  5G Wireless Systems}.\hskip 1em plus 0.5em minus 0.4em\relax Cambridge
  University Press, 2017.

\bibitem{spectrumreport}
``Spectrum policy task force report,'' Federal Commun. Comm., Washington, DC,
  ET Docket No. 02-135, Nov. 2002.

\bibitem{datla2009spectrum}
D.~Datla, A.~M. Wyglinski, and G.~J. Minden, ``A spectrum surveying framework
  for dynamic spectrum access networks,'' \emph{IEEE Trans. Veh. Tech.},
  vol.~58, no.~8, pp. 4158--4168, Oct. 2009.

\bibitem{4570202}
H.~{Islam}, Y.~{Liang}, and A.~T. {Hoang}, ``Joint power control and
  beamforming for cognitive radio networks,'' \emph{IEEE Trans. Wireless
  Commun.}, vol.~7, no.~7, pp. 2415--2419, Jul. 2008.

\bibitem{7111366}
D.~W.~K. {Ng}, E.~S. {Lo}, and R.~{Schober}, ``Multiobjective resource
  allocation for secure communication in cognitive radio networks with wireless
  information and power transfer,'' \emph{IEEE Trans. Veh. Tech.}, vol.~65,
  no.~5, pp. 3166--3184, May 2016.

\bibitem{6832464}
A.~{Sabharwal}, P.~{Schniter}, D.~{Guo}, D.~W. {Bliss}, S.~{Rangarajan}, and
  R.~{Wichman}, ``In-band full-duplex wireless: challenges and opportunities,''
  \emph{IEEE J. Select. Areas Commun.}, vol.~32, no.~9, pp. 1637--1652, Sep.
  2014.

\bibitem{afifi2014incorporating}
W.~Afifi and M.~Krunz, ``Incorporating self-interference suppression for
  full-duplex operation in opportunistic spectrum access systems,'' \emph{IEEE
  Trans. Wireless Commun.}, vol.~14, no.~4, pp. 2180--2191, Apr. 2014.

\bibitem{7842235}
S.~{Xiao}, X.~{Zhou}, G.~Y. {Li}, and W.~{Guo}, ``Robust resource allocation in
  full-duplex cognitive radio networks,'' in \emph{Proc. IEEE Global Commun.
  Conf. (GLOBECOM)}, Washington, DC, USA, Dec. 2016, pp. 1--7.

\bibitem{7760353}
Y.~Sun, D.~W.~K. Ng, N.~Zlatanov, and R.~Schober, ``Robust resource allocation
  for full-duplex cognitive radio systems,'' in \emph{Proc. 24th European
  Signal Processing Conference (EUSIPCO)}, Budapest, Hungary, Aug. 2016, pp.
  773--777.

\bibitem{6493535}
G.~{Zheng}, I.~{Krikidis}, and B.~{Ottersten}, ``Full-duplex cooperative
  cognitive radio with transmit imperfections,'' \emph{IEEE Trans. Wireless
  Commun.}, vol.~12, no.~5, pp. 2498--2511, May 2013.

\bibitem{di2019smart}
M.~Di~Renzo \emph{et~al.}, ``Smart radio environments empowered by
  reconfigurable {AI} meta-surfaces: an idea whose time has come,''
  \emph{EURASIP Journal on Wireless Communications and Networking}, vol. 2019,
  no.~1, pp. 1--20, May 2019.

\bibitem{wu2019intelligent}
Q.~Wu and R.~Zhang, ``Intelligent reflecting surface enhanced wireless network
  via joint active and passive beamforming,'' \emph{IEEE Trans. Wireless
  Commun.}, vol.~18, no.~11, pp. 5394--5409, Aug. 2019.

\bibitem{yu2019miso}
X.~Yu, D.~Xu, and R.~Schober, ``{MISO} wireless communication systems via
  intelligent reflecting surfaces,'' in \emph{Proc. IEEE Int. Conf. Commun.
  China (ICCC)}, Changchun, China, May 2019, pp. 1--6.

\bibitem{xu2019resource}
D.~Xu, X.~Yu, Y.~Sun, D.~W.~K. Ng, and R.~Schober, ``Resource allocation for
  secure {IRS}-assisted multiuser {MISO} systems,'' in \emph{Proc. IEEE Global
  Commun. Conf. (GLOBECOM) Workshops}, Waikoloa, HI, USA, Dec. 2019, pp. 1--6.

\bibitem{pan2019intelligent}
C.~Pan, H.~Ren, K.~Wang, M.~Elkashlan, A.~Nallanathan, J.~Wang, and L.~Hanzo,
  ``Intelligent reflecting surface enhanced {MIMO} broadcasting for
  simultaneous wireless information and power transfer,'' \emph{arXiv preprint
  arXiv:1908.04863}, 2019.

\bibitem{zhang2019capacity}
S.~Zhang and R.~Zhang, ``Capacity characterization for intelligent reflecting
  surface aided {MIMO} communication,'' \emph{arXiv preprint arXiv:1910.01573},
  2019.

\bibitem{zhang2019multiple}
J.~Zhang, E.~Bj{\"o}rnson, M.~Matthaiou, D.~W.~K. Ng, H.~Yang, and D.~J. Love,
  ``Multiple antenna technologies for beyond 5{G},'' \emph{arXiv preprint
  arXiv:1910.00092}, 2019.

\bibitem{8910627}
Q.~{Wu} and R.~{Zhang}, ``Towards smart and reconfigurable environment:
  Intelligent reflecting surface aided wireless network,'' \emph{IEEE Commun.
  Mag.}, vol.~58, no.~1, pp. 106--112, Jan. 2020.

\bibitem{cui2014coding}
T.~J. Cui, M.~Q. Qi, X.~Wan, J.~Zhao, and Q.~Cheng, ``Coding metamaterials,
  digital metamaterials and programmable metamaterials,'' \emph{Light: Science
  \& Applications}, vol.~3, no.~10, p. e218, 2014.

\bibitem{yang2019irs}
Y.~{Yang}, S.~{Zhang}, and R.~{Zhang}, ``{IRS}-enhanced {OFDM}: Power
  allocation and passive array optimization,'' \emph{arXiv preprint
  arXiv:1905.00604}, 2019.

\bibitem{yu2019enabling}
X.~Yu, D.~Xu, and R.~Schober, ``Enabling secure wireless communications via
  intelligent reflecting surfaces,'' in \emph{Proc. IEEE Global Commun. Conf.
  (GLOBECOM), Waikoloa, HI, USA}, Dec. 2019, pp. 1--6.

\bibitem{tseng2001convergence}
P.~Tseng, ``Convergence of a block coordinate descent method for
  nondifferentiable minimization,'' \emph{Journal of Optimization Theory and
  Applications}, vol. 109, no.~3, pp. 475--494, Jun. 2001.

\bibitem{dinh2010local}
Q.~T. Dinh and M.~Diehl, ``Local convergence of sequential convex programming
  for nonconvex optimization,'' in \emph{Recent Advances in Optimization and
  its Applications in Engineering}.\hskip 1em plus 0.5em minus 0.4em\relax
  Springer, 2010.

\bibitem{ben1997penalty}
A.~Ben-Tal and M.~Zibulevsky, ``Penalty/barrier multiplier methods for convex
  programming problems,'' \emph{SIAM Journal on Optimization}, vol.~7, no.~2,
  pp. 347--366, 1997.

\bibitem{liu2016nonorthogonal}
Y.~Liu, Z.~Ding, M.~Elkashlan, and J.~Yuan, ``Nonorthogonal multiple access in
  large-scale underlay cognitive radio networks,'' \emph{IEEE Trans. Veh.
  Tech.}, vol.~65, no.~12, pp. 10\,152--10\,157, 2016.

\bibitem{bharadia2013full}
D.~Bharadia, E.~McMilin, and S.~Katti, ``Full duplex radios,'' in \emph{ACM
  SIGCOMM}, vol.~43, no.~4, 2013, pp. 375--386.

\bibitem{ghosh2010fundamentals}
A.~Ghosh, J.~Zhang, J.~G. Andrews, and R.~Muhamed, \emph{Fundamentals of
  LTE}.\hskip 1em plus 0.5em minus 0.4em\relax Pearson Education, 2010.

\bibitem{zhang2009robust}
L.~Zhang, Y.-C. Liang, Y.~Xin, and H.~V. Poor, ``Robust cognitive beamforming
  with partial channel state information,'' \emph{IEEE Trans. Wireless
  Commun.}, vol.~8, no.~8, pp. 4143--4153, Aug. 2009.

\bibitem{wang2019channel}
Z.~Wang, L.~Liu, and S.~Cui, ``Channel estimation for intelligent reflecting
  surface assisted multiuser communications,'' \emph{arXiv preprint
  arXiv:1911.03084}, 2019.

\bibitem{wang2009worst}
J.~Wang and D.~P. Palomar, ``Worst-case robust {MIMO} transmission with
  imperfect channel knowledge,'' \emph{IEEE Trans. Signal Process.}, vol.~57,
  no.~8, pp. 3086--3100, Aug. 2009.

\bibitem{day2012full}
B.~P. Day, A.~R. Margetts, D.~W. Bliss, and P.~Schniter, ``Full-duplex {MIMO}
  relaying: Achievable rates under limited dynamic range,'' in \emph{Proc.
  Forty Sixth Asilomar Conf. Signals, Systems and Computers (ASILOMAR)}, 2012,
  pp. 1290--1294.

\bibitem{7463025}
Y.~Sun, D.~W.~K. Ng, J.~Zhu, and R.~Schober, ``Multi-objective optimization for
  robust power efficient and secure full-duplex wireless communication
  systems,'' \emph{IEEE Trans. Wireless Commun.}, vol.~15, no.~8, pp.
  5511--5526, Apr. 2016.

\bibitem{masmoudi2016channel}
A.~Masmoudi and T.~Le-Ngoc, ``Channel estimation and self-interference
  cancelation in full-duplex communication systems,'' \emph{IEEE Trans. Veh.
  Tech.}, vol.~66, no.~1, pp. 321--334, Jan. 2016.

\bibitem{li2013safe}
Q.~Li, W.-K. Ma, and A.~M.-C. So, ``A safe approximation approach to secrecy
  outage design for {MIMO} wiretap channels,'' \emph{IEEE Signal Process.
  Lett.}, vol.~21, no.~1, pp. 118--121, Dec. 2013.

\bibitem{boyd2004convex}
S.~Boyd and L.~Vandenberghe, \emph{Convex Optimization}.\hskip 1em plus 0.5em
  minus 0.4em\relax Cambridge University Press, 2004.

\bibitem{grant2008cvx}
M.~Grant and S.~Boyd, ``{CVX}: Matlab software for disciplined convex
  programming, version 2.1,'' \emph{Available at http://cvxr.com/cvx}, Mar.
  2017.

\bibitem{gershman2010convex}
A.~B. Gershman, N.~D. Sidiropoulos, S.~Shahbazpanahi, M.~Bengtsson, and
  B.~Ottersten, ``Convex optimization-based beamforming,'' \emph{IEEE Signal
  Process. Mag.}, vol.~27, no.~3, pp. 62--75, May 2010.

\bibitem{5447068}
Z.~{Luo}, W.~{Ma}, A.~M. {So}, Y.~{Ye}, and S.~{Zhang}, ``Semidefinite
  relaxation of quadratic optimization problems,'' \emph{IEEE Signal Process.
  Mag.}, vol.~27, no.~3, pp. 20--34, May 2010.

\bibitem{8761429}
K.~{Yang}, T.~{Jiang}, Y.~{Shi}, and Z.~{Ding}, ``Federated learning based on
  over-the-air computation,'' in \emph{Proc. Intern. Conf. Commun. (ICC)},
  Shanghai, China, May 2019, pp. 1--6.

\bibitem{polik2010interior}
I.~P{\'o}lik and T.~Terlaky, ``Interior point methods for nonlinear
  optimization,'' in \emph{Nonlinear Optimization}.\hskip 1em plus 0.5em minus
  0.4em\relax Springer, 2010, pp. 215--276.

\end{thebibliography}
\end{document}